\documentstyle[epsf,12pt]{article}

\begin{document}

\title
{Two-loop renormalization of $N=1$ supersymmetric electrodynamics,
regularized by higher derivatives.}

\author{A.Soloshenko\thanks{E-mail:$solosh@theor.phys.msu.su$} and
K.Stepanyantz \thanks{E-mail:$stepan@theor.phys.msu.su$}}

\maketitle

\begin{center}
{\em Moscow State University, physical faculty,\\
department of theoretical physics.\\
$117234$, Moscow, Russia}
\end{center}

\begin{abstract}
Two-loop $\beta$-function and anomalous dimension are calculated for
$N=1$ supersymmetric quantum electrodynamics, regularized by higher
derivatives in the minimal subtraction scheme. The result for two-loop
contribution to the $\beta$-function appears to be equal to 0, does not
depend on the form of regularizing term and does not lead to anomaly
puzzle. Two-loop anomalous dimension can be also made independent on
parameters of higher derivative regularization by a special choice of
subtraction scheme.
\end{abstract}

\sloppy

%%%%%%%%%%%%%%%%%%%%%%%%%%%%%%%%%%%%%%%%%%%%%%%%%%%%%%%%%%%%%%%%%%%%%%%%%%

\section{Introduction.}
\hspace{\parindent}

Investigation of quantum corrections in supersymmetric theories is
a very important and complicated problem. In principle, supersymmetric
theories have better ultraviolet behavior, than nonsupersymmetric
models. For example, in $N=2$ Yang-Mills theory perturbative divergences
are present only in one-loop diagrams. In principle it follows from
the fact, that in supersymmetric theories axial anomaly and anomaly
of energy-momentum tensor trace belong to the same supermultiplet
\cite{Ferrara,Clark,Piquet1,Piquet2}. The axial anomaly is known to
be completely defined by the one-loop approximation
\cite{Bardeen,Slavnov_Book}, while the trace anomaly is proportional
to $\beta$-function \cite{Adler_Collins}. Therefore, due to the
supersymmetry the $\beta$-function should be also defined by the one-loop
approximation. The same arguments can be applied to $N=1$ supersymmetric
theories. However, explicit calculations of radiative corrections show,
that the $\beta$-function in $N=1$ supersymmetric models has contributions
from higher loops \cite{Tarasov,Grisaru,Caswell,Avdeev}. This contradiction
is usually called "anomaly puzzle" and was investigated in a large number
of papers, for example \cite{12,13,14,15,16,18,19,20,21,22,NSVZ,Arkani}.

Usually different proposals to solve anomaly puzzle require to fix the
form of the $\beta$-function in all orders of perturbation theory. For
example, in theories with matter the $\beta$-function should be related
with the anomalous dimension. For the first time such $\beta$-function
was obtained by Novikov, Shifman, Vainshtein and Zakharov (NSVZ) from
investigation of structure of instanton corrections \cite{NSVZ_Instanton}.
Later this result was checked by explicit calculations, which were
usually made by the dimensional reduction technique \cite{Siegel}.
Two-loop $\beta$-function, obtained in this regularization, is shown to
coincide with a prediction, following from NSVZ exact expression. However,
three-loop $\beta$-function \cite{ThreeLoop1,ThreeLoop2,ThreeLoop3} does
not agree with it. Nevertheless, the deviations can be removed by a
redefinition of the coupling constant \cite{Jones}, the possibility of
such redefinition being highly nontrivial \cite{JackJones}. In principle
it is possible to relate $\overline{\mbox{DR}}$ scheme and NSVZ scheme
order by order \cite{North} in the perturbation theory.

It is necessary to especially mention paper \cite{Arkani}, in which
$\beta$-function is shown to depend on the normalization of matter
and gauge superfields. In particular, NSVZ $\beta$-function can be
obtained after a special rescaling of these superfields, which reduces
kinetic terms to the canonical normalization. Otherwise, (without
rescalings) $\beta$-function is argued to be completely defined by
the one-loop approximation. So, it is really quite possible to obtain
zero contributions of higher loops. The problem is how to calculate
the corrections. In principle, it is possible to look for the
regularization or renormalization scheme, in which $\beta$-function is
equal to the one-loop result or coincides with NSVZ exact $\beta$-function.
For example, the calculation of super Yang-Mills two-loop $\beta$-function
in the differential renormalization \cite{DiffR} was made in \cite{Mas}.
Another interesting possibility is using of higher covariant derivative
regularization \cite{Slavnov,Bakeyev}. For the supersymmetric Yang-Mills
theory the Lagrangian of the regularized theory was constructed in
\cite{West_Paper}. For electrodynamics construction of the regularized
Lagrangian is simpler, because instead of covariant derivatives it is
necessary to use usual derivatives. However, calculation of diagrams,
regularized by higher covariant derivatives is rather complicated. In
particular, explicit calculation of the one-loop quantum correction
\footnote{Note, that introducing of a term with higher covariant
derivatives does not lead to regularization of one-loop divergences. For
the one-loop divergences it is necessary to use one more regularization,
for example, introduce Pauli-Villars fields.} for the (nonsupersymmetric)
Yang-Mills theory was made rather recently \cite{Martin,Asorey} and gives
the same result as the dimensional regularization. In principle it is
possible to prove, that one-loop calculations using higher covariant
derivative regularization (certainly, complemented by the additional
regularization for one-loop diagrams) always give the same result as
the dimensional regularization \cite{PhysLett}. Investigation of two-loop
corrections in theories, regularized by higher derivatives has not yet
been done.

In this paper we try to understand features of higher derivative
regularization in supersymmetric theories and calculate two-loop
renormgroup functions for massless $N=1$ supersymmetric quantum
electrodynamics in this regularization using minimal subtraction scheme.

The paper is organized as follows:

In Section \ref{Section_SUSY_QED} we introduce notations and remind
some information about $N=1$ supersymmetric electrodynamics. In the
next Section \ref{Section_Covariant_Derivatives} the considered model
is regularized by adding of higher derivative term. After it we describe
the quantization procedure for the constructed theory. Two-loop
$\beta$-function and anomalous dimension are calculated in Section
\ref{Section_Two_Loop}. Agreement of the results with renormgroup
equations is checked in Section \ref{Section_Renormgroup}. A brief
summary and discussion are presented in Conclusion. Technical details
of calculations, including expressions for all Feinman diagrams, can
be found in the Appendix.

%%%%%%%%%%%%%%%%%%%%%%%%%%%%%%%%%%%%%%%%%%%%%%%%%%%%%%%%%%%%%%%%%%%%%%%%%

\section{Supersymmetric quantum electrodynamics.}
\hspace{\parindent}
\label{Section_SUSY_QED}

$N=1$ supersymmetric massless electrodynamics in the superspace is
described by the following action:
\footnote{In our notations the metric tensor in the Minkowski
space-time has the diagonal elements (1, -1, -1, -1).}

\begin{equation}\label{SQED_Action}
S_0 = \frac{1}{4 e^2} \mbox{Re}\int d^4x\,d^2\theta\,W_a C^{ab} W_b
+ \frac{1}{4}\int d^4x\, d^4\theta\,
\Big(\phi^* e^{2V}\phi +\tilde\phi^*
e^{-2V}\tilde\phi\Big).
\end{equation}

\noindent
Here $\phi$ and $\tilde\phi$ are chiral superfields, which
in components can be written as

\begin{eqnarray}\label{Phi_Superfield}
&& \phi(y,\theta) = \varphi(y) + \bar\theta (1+\gamma_5) \psi(y)
+ \frac{1}{2}\bar\theta (1+\gamma_5)\theta f(y);\nonumber\\
&&\tilde \phi(y,\theta) = \tilde \varphi(y)
+ \bar\theta (1+\gamma_5) \tilde \psi(y)
+ \frac{1}{2}\bar\theta (1+\gamma_5)\theta \tilde f(y),
\end{eqnarray}

\noindent
where $y^\mu = x^\mu + i\bar\theta\gamma^\mu\gamma_5\theta/2$ are chiral
coordinates, $\varphi$ and $\tilde\varphi$ are complex scalar fields,
$\psi$ and $\tilde\psi$ are Maiorana spinors, which can be unified
in a Dirac spinor

\begin{equation}
\Psi = \frac{1}{\sqrt{2}}\Big((1+\gamma_5)\psi+(1-\gamma_5)\tilde\psi\Big),
\end{equation}

\noindent
and $f$ and $\tilde f$ are auxiliary complex scalar fields.

$V$ is a real abelian superfield, which is a supersymmetric
generalization of the gauge field. In components this superfield
can be written as

\begin{eqnarray}\label{V_Superfield}
&& V(x,\theta) = C(x)+i\sqrt{2}\bar\theta\gamma_5\xi(x)
+\frac{1}{2}(\bar\theta\theta)K(x)
+\frac{i}{2}(\bar\theta\gamma_5\theta)H(x)
+\frac{1}{2}(\bar\theta \gamma^\mu \gamma_5\theta) A_\mu(x)
+\nonumber\\
&& + \sqrt{2} (\bar\theta\theta) \bar\theta
\Big(i\gamma_5\chi(x)
+\frac{1}{2}\gamma^\mu\gamma_5\partial_\mu\xi(x)\Big)
+ \frac{1}{4} (\bar\theta\theta)^2 \Big(D(x)
-\frac{1}{2}\partial^2 C(x)\Big).
\end{eqnarray}

The chiral superfield $W_a$ is a supersymmetric generalization of
the field strength tensor and in the abelian case is defined as

\begin{equation}\label{W_Superfield}
W_a = \frac{1}{16} \bar D (1-\gamma_5) D\Big[(1+\gamma_5)D_a V\Big],
\end{equation}

\noindent
where the supersymmetric covariant derivative $D$ is written as

\begin{equation}
D = \frac{\partial}{\partial\bar\theta} - i\gamma^\mu\theta\,\partial_\mu.
\end{equation}

Model (\ref{SQED_Action}) is invariant under supersymmetric
gauge transformations

\begin{equation}\label{Gauge_Transformations}
V \to V - \frac{1}{2}(A+A^+);
\qquad \phi\to e^{A}\phi;\qquad \tilde\phi\to e^{-A} \tilde\phi,
\end{equation}

\noindent
where $A$ is an arbitrary chiral superfield. In principle, it is
possible to choose Wess-Zumino gauge, in which the superfield $V$
is written as

\begin{eqnarray}
V(x,\theta) =
\frac{1}{2}(\bar\theta \gamma^\mu \gamma_5\theta) A_\mu(x)
+ i\sqrt{2} (\bar\theta\theta) \bar\theta\gamma_5\chi(x)
+ \frac{1}{4} (\bar\theta\theta)^2 D(x).
\end{eqnarray}

\noindent
However, this gauge is not supersymmetric. That is why we do not
use it for calculation of quantum corrections.

%%%%%%%%%%%%%%%%%%%%%%%%%%%%%%%%%%%%%%%%%%%%%%%%%%%%%%%%%%%%%%%%%%%%%%%%%%%

\section{Higher derivative regularization of $N=1$ supersymmetric
electrodynamics.}
\hspace{\parindent}
\label{Section_Covariant_Derivatives}

To regularize model (\ref{SQED_Action}) by higher derivatives let us
first modify its action by the following way:

\begin{eqnarray}\label{Regularized_SQED_Action}
&& S_0 \to S = S_0 + S_{\Lambda}
=\vphantom{\frac{1}{2}}\nonumber\\
&&\qquad
= \frac{1}{4 e^2} \mbox{Re}\int d^4x\,d^2\theta\,W_a C^{ab}
\Big(1 + \frac{\partial^{2n}}{\Lambda^{2n}}\Big) W_b
+\nonumber\\
&&\qquad\qquad\qquad\qquad\qquad\qquad
+ \frac{1}{4}\int d^4x\, d^4\theta\,
\Big(\phi^* e^{2V}\phi +\tilde\phi^* e^{-2V}\tilde\phi\Big).\qquad
\end{eqnarray}

Note, that the considered model is abelian and the superfield
$W^a$ is gauge invariant. Therefore, a regularizing term should
contain usual derivatives instead of the covariant ones.

It is convenient to introduce operators

\begin{eqnarray}
&& \bar D^2 \equiv \frac{1}{2} \bar D (1-\gamma_5) D;\qquad
D^2 \equiv \frac{1}{2} \bar D (1+\gamma_5) D;\nonumber\\
&&\qquad
\Pi_{1/2} \equiv - \frac{1}{16\partial^2} D_a \Big(C(1+\gamma_5)\Big)^{ab}
\bar D^2 D_b,
\end{eqnarray}

\noindent
which satisfy identities

\begin{eqnarray}\label{Identity1}
&& D^2 \bar D^2 + \bar D^2 D^2
= -16\Pi_{1/2}\partial^2 - 16\partial^2;\\
\label{Identity2}
&& \frac{1}{2} \bar D \gamma^{\mu} \gamma_5 D\,
\bar D \gamma^{\nu} \gamma_5 D
+ \frac{1}{2} \bar D \gamma^{\nu} \gamma_5 D\,
\bar D \gamma^{\mu} \gamma_5 D
= - 16\eta^{\mu\nu}\Pi_{1/2} \partial^2 - 16 \partial^\mu \partial^\nu.
\qquad
\end{eqnarray}

\noindent
Then the first term in action (\ref{Regularized_SQED_Action}) can be
presented in the following form:

\begin{eqnarray}
&& S_{\mbox{\footnotesize gauge}}\equiv
\frac{1}{4e^2}\mbox{Re}\int d^4x\, d^2\theta\,
W_a C^{ab} \Big(1 + \frac{\partial^{2n}}{\Lambda^{2n}}\Big)W_b
=\nonumber\\
&& \qquad\qquad\qquad\qquad\qquad\qquad
= - \frac{1}{4e^2} \int d^4x\, d^4\theta\, V \Pi_{1/2}\partial^2
\Big(1 + \frac{\partial^{2n}}{\Lambda^{2n}}\Big) V.\qquad\quad
\end{eqnarray}

The gauge invariance (\ref{Gauge_Transformations}) can be fixed by
addition of the terms

\begin{equation}
S_{\mbox{\footnotesize gf}} = - \frac{1}{64 e^2}\int d^4x\,d^4\theta\,
\Bigg(V D^2 \bar D^2
\Big(1 + \frac{\partial^{2n}}{\Lambda^{2n}}\Big) V
+ V \bar D^2 D^2
\Big(1+ \frac{\partial^{2n}}{\Lambda^{2n}}\Big) V\Bigg),
\end{equation}

\noindent
which are invariant under supersymmetry transformations. Then due
to identity (\ref{Identity1}) the kinetic term for the gauge field
is written in the most simple form:

\begin{equation}
S_{\mbox{\footnotesize gauge}}
+ S_{\mbox{\footnotesize gf}} = \frac{1}{4 e^2}\int d^4x\,d^4\theta
V\partial^2 \Big(1+ \frac{\partial^{2n}}{\Lambda^{2n}}\Big) V.
\end{equation}

Due to the gauge invariance (\ref{Gauge_Transformations}) the
renormalized action of the considered model (without gauge fixing
term) can be presented as

\begin{eqnarray}\label{Two_Loop_S_Ren}
&& S_{\mbox{\footnotesize ren}} =
\frac{1}{4 e^2} Z_3(\Lambda/\mu)\,
\mbox{Re}\int d^4x\,d^2\theta\,W_a C^{ab}
\Big(1+ \frac{\partial^{2n}}{\Lambda^{2n}}\Big) W_b
+\nonumber\\
&& \qquad\qquad\qquad\qquad\qquad
+ \frac{1}{4} Z(\Lambda/\mu) \int d^4x\, d^4\theta\,
\Big(\phi^* e^{2V}\phi +\tilde\phi^* e^{-2V}\tilde\phi\Big),\qquad
\end{eqnarray}

\noindent
where $e=e(\Lambda/\mu)$ is a renormalized coupling constant.
A bare coupling constant $e_0$ is defined by the equation

\begin{equation}
\frac{1}{e_0^2} = \frac{1}{e^2} Z_3(\Lambda/\mu)
\end{equation}

\noindent
and does not depend on $\Lambda/\mu$. The $\beta$-function and anomalous
dimension in our notations are defined as

\begin{equation}
\beta = \frac{d}{d\ln\mu}\Bigg(\frac{e^2}{4\pi}\Bigg);
\qquad\quad
\gamma = \frac{d\ln Z}{d\ln\mu}.
\end{equation}

At the first sight, the generating functional can be written in the
following form:

\begin{eqnarray}\label{Z}
&& Z = \int DV\,D\phi\,D\tilde \phi\,
\exp\Bigg\{i\Bigg[\frac{1}{4 e^2} \int d^4x\,d^4\theta\, V\partial^2
\Big(1+ \frac{\partial^{2n}}{\Lambda^{2n}}\Big) V
-\nonumber\\
&& \qquad\qquad\qquad\qquad
- \frac{1}{4 e^2} \Big(Z_3(\Lambda/\mu)-1\Big)
\int d^4x\,d^4\theta\, V \Pi_{1/2}\partial^2
\Big(1+ \frac{\partial^{2n}}{\Lambda^{2n}}\Big) V
+\nonumber\\
&& + \frac{1}{4} Z(\Lambda/\mu) \int d^4x\,d^4\theta\,
\Big(\phi^* e^{2V}\phi
+ \tilde\phi^* e^{-2V}\tilde\phi \Big)
+ \int d^4x\,d^4\theta\,J V
+\nonumber\\
&& \qquad\qquad\qquad\qquad
+ \int d^4x\,d^2\theta\, \Big(j\,\phi + \tilde j\,\tilde\phi \Big)
+ \int d^4x\,d^2\bar\theta\,
\Big(j^*\phi^* + \tilde j^* \tilde\phi^* \Big)\Bigg]\Bigg\}.\qquad\quad
\end{eqnarray}

\noindent
(Below we slightly modify this expression.) Note, that the considered
case corresponds to the gauge group $U(1)$ and, therefore, diagrams
with ghost loops are absent.

Taking into account, that

\begin{equation}\label{Chiral_Identity}
\bar D^2 D^2 \phi = - 16\partial^2\phi
\end{equation}

\noindent
for any chiral superfield $\phi$ and that

\begin{equation}
\int d^4x\,d^2\theta = -\frac{1}{2} \int d^4x\,D^2,
\end{equation}

\noindent
generating functional (\ref{Z}) can be presented as

\begin{eqnarray}
&& Z = \int DV\,D\phi\,D\tilde \phi\,
\exp\Bigg\{i \int d^4x\,d^4\theta\,
\Bigg(\frac{1}{4 e^2} V \partial^2
\Big(1+ \frac{\partial^{2n}}{\Lambda^{2n}}\Big) V
-\nonumber\\
&& - \frac{1}{4 e^2} \Big(Z_3(\Lambda/\mu)-1\Big)\,
V \Pi_{1/2}\partial^2
\Big(1+ \frac{\partial^{2n}}{\Lambda^{2n}}\Big) V
+ \frac{1}{4} Z(\Lambda/\mu)\Big(\phi^* e^{2V}\phi
+ \tilde\phi^* e^{-2V}\tilde\phi\Big)
+\nonumber\\
&& \qquad\qquad\qquad\qquad\qquad\qquad
+ J V +  \phi \frac{D^2}{8\partial^2} j
+ \tilde\phi \frac{D^2}{8\partial^2}\tilde j
+ \phi^* \frac{\bar D^2}{8\partial^2} j^*
+  \tilde\phi^* \frac{\bar D^2}{8\partial^2} \tilde j^*\Bigg) \Bigg\}.
\qquad\nonumber\\
\end{eqnarray}

In order to calculate this functional we should present the argument
of the exponent as a sum of a part $S_Q$, quadratic in fields, and
interaction $S_I$, where

\begin{eqnarray}\label{Quadratic}
&& S_Q = \frac{1}{4 e^2} \int d^4x\,d^4\theta\,\Bigg(V\partial^2
\Big(1 + \frac{\partial^{2n}}{\Lambda^{2n}}\Big) V
+ \frac{1}{4} \Big(\phi^* \phi +\tilde\phi^* \tilde\phi\Big)
+\nonumber\\
&& \qquad\qquad\qquad\qquad
+ J V +  \phi \frac{D^2}{8\partial^2} j
+ \tilde\phi \frac{D^2}{8\partial^2}\tilde j
+ \phi^* \frac{\bar D^2}{8\partial^2} j^*
+  \tilde\phi^* \frac{\bar D^2}{8\partial^2} \tilde j^* \Bigg);
\nonumber\\
\label{Interaction}
&& S_I = \sum\limits_{n=1}^\infty \frac{1}{4 n!}\int d^4x\,d^4\theta\,
\Big(\phi^* (2V)^{n} \phi + \tilde\phi^* (-2V)^n\tilde\phi\Big)
-\nonumber\\
&& \qquad\qquad
- \Big(Z_3(\Lambda/\mu)-1\Big) \frac{1}{4 e^2}\int d^4x\,d^4\theta\,
V\Pi_{1/2} \partial^2 \Big(1+\frac{\partial^{2n}}{\Lambda^{2n}}\Big) V
+\nonumber\\
&& \qquad\qquad\qquad\qquad
+ \Big(Z(\Lambda/\mu) - 1\Big) \frac{1}{4}\int d^4x\,d^4\theta\,
\Big(\phi^* e^{2V} \phi + \tilde\phi^* e^{-2V}\tilde\phi\Big).\qquad\quad
\vphantom{\Bigg(}
\end{eqnarray}

\noindent
(Last two terms of $S_I$ will generate vertexes with insersions of
counterterms, obtained in previous orders of perturbation theory.)

Then the generating functional can be written as

\begin{eqnarray}\label{Z_With_SI}
&& Z = \exp\Bigg\{i S_I\Bigg(\frac{1}{i}\frac{\delta}{\delta J},\,
\frac{1}{i}\frac{\delta}{\delta j},\ldots\Bigg)\Bigg\}
\int DV\,D\phi\,D\tilde \phi\,\exp\Big(i S_Q\Big).
\end{eqnarray}

\noindent
Note, that the differentiation over chiral superfields in our notations
is defined as follows:

\begin{equation}
\frac{\delta j(\theta_x,x)}{\delta j(\theta_y,y)} = - \frac{\bar D^2}{2}
\delta^4(\theta_x-\theta_y)\,\delta^4(x-y),
\end{equation}

\noindent
so that

\begin{equation}
\frac{\delta}{\delta j(\theta_y,y)}
\int d^4x\,d^2\theta_x\,j(x,\theta_x)\,\phi(x,\theta_x)
= \phi(y,\theta_y).
\end{equation}

\noindent
The integral, remaining in equation (\ref{Z_With_SI}), is Gaussian and
can be easily calculated:

\begin{eqnarray}\label{Perturbative_Z }
&& Z = \exp\Bigg\{i S_I\Bigg(\frac{1}{i}\frac{\delta}{\delta J},\,
\frac{1}{i}\frac{\delta}{\delta j},\ldots\Bigg)\Bigg\}
\times\nonumber\\
&& \qquad\qquad\quad
\times
\exp\Bigg\{ i \int d^4x\,d^4\theta\,\Bigg(
j^* \frac{1}{\partial^2}  j
+ \tilde j^* \frac{1}{\partial^2} \tilde j
- J \frac{e^2}{\partial^2
\Big(1+\partial^{2n}/\Lambda^{2n}\Big)} J\Bigg)\Bigg\}\qquad
\end{eqnarray}

\noindent
Expansion of this expression in powers of $J$, $j$ and $\tilde j$ gives
a series of perturbation theory.

However, introducing of higher derivative term does not eliminate all
divergences. Really, in Appendix \ref{Appendix_Degree_Of_Divergence}
we check, that the superficial degree of divergence for model
(\ref{Regularized_SQED_Action}) is equal to

\begin{equation}\label{Degree_Of_Divergence}
\omega_\Lambda = 2 - 2n (L-1) - E_\phi (n+1),
\end{equation}

\noindent
where $L$ is a number of loops and $E_\phi$ is a number of external
$\phi$-lines. Note, that $\omega_\Lambda$ does not depend on a number
of external $V$-lines $E_V$. Therefore even after introducing of the
higher derivative term with $n\ge 2$, divergences are present in one-loop
diagrams. In order to regularize them \cite{Slavnov_Book} it is necessary
to insert Pauli-Villars determinants in generating functional (\ref{Z}),
so that

\begin{eqnarray}\label{Modified_Z}
&& Z = \int DV\,D\phi\,D\tilde \phi\,
\prod\limits_i \Big(\det PV(V,M_i)\Big)^{c_i}
\exp\Bigg\{i\Bigg[\frac{1}{4 e^2} \int d^4x\,d^4\theta\, V\partial^2
\Big(1+ \frac{\partial^{2n}}{\Lambda^{2n}}\Big) V
-\nonumber\\
&& - \frac{1}{4 e^2} \Big(Z_3(\Lambda/\mu)-1\Big) \int d^4x\,d^4\theta\,
V \Pi_{1/2}\partial^2
\Big(1+ \frac{\partial^{2n}}{\Lambda^{2n}}\Big) V
+\nonumber\\
&& \qquad\qquad\qquad\qquad\qquad\qquad\qquad\qquad
+ \frac{1}{4} Z(\Lambda/\mu) \int d^4x\,d^4\theta\,
\Big(\phi^* e^{2V}\phi
+ \tilde\phi^* e^{-2V}\tilde\phi \Big)
+\nonumber\\
&&
+ \int d^4x\,d^4\theta\,J V
+ \int d^4x\,d^2\theta\, \Big(j\,\phi + \tilde j\,\tilde\phi \Big)
+ \int d^4x\,d^2\bar\theta\,
\Big(j^*\phi^* + \tilde j^* \tilde\phi^* \Big)\Bigg]\Bigg\},
\end{eqnarray}

\noindent
where

\begin{eqnarray}
&& \Big(\det PV(V,M)\Big)^{-1} = \int D\Phi\,D\tilde \Phi\,
\exp\Bigg\{i\Bigg[ Z(\Lambda/\mu) \frac{1}{4} \int d^4x\,d^4\theta\,
\Big(\Phi^* e^{2V}\Phi
+\qquad\nonumber\\
&& + \tilde\Phi^* e^{-2V}\tilde\Phi \Big)
+ \frac{1}{2}\int d^4x\,d^2\theta\, M \tilde\Phi \Phi
+ \frac{1}{2}\int d^4x\,d^2\bar\theta\, M \tilde\Phi^* \Phi^*
\Bigg]\Bigg\}
\end{eqnarray}

\noindent
and the coefficients $c_i$ satisfy equations

\begin{equation}
\sum\limits_i c_i = 1;\qquad \sum\limits_i c_i M_i^2 = 0.
\end{equation}

\noindent
Below we will also assume, that $M_i = a_i\Lambda$, where $a_i$
are constants. Insersion of such Pauli-Villars determinants allows
to cancel remaining divergences in all one-loop diagrams, including
diagrams with insersions of counterterms.

Repeating the above arguments it is possible to write the Pauli-Villars
determinants in the following form

\begin{eqnarray}
&& \Big(\det PV(V,M)\Big)^{-1}
= \exp\Bigg\{i (S_{PV})_I\Bigg(V,\,
\frac{1}{i}\frac{\delta}{\delta {\bf j}},\ldots\Bigg)\Bigg\}
\exp\Bigg\{ i \int d^4x\,d^4\theta\,
\times
\nonumber\\
&& \times
\Bigg(
{\bf j}^* \frac{1}{\partial^2+M^2}  {\bf j}
+ \tilde {\bf j}^* \frac{1}{\partial^2+M^2} \tilde {\bf j}
+ {\bf j} \frac{M}{\partial^2+M^2}\frac{D^2}{4\partial^2} \tilde {\bf j}
+ {\bf j}^* \frac{M}{\partial^2+M^2} \frac{\bar D^2}{4\partial^2}
\tilde {\bf j}^*
\Bigg)\Bigg\}\Bigg|_{{\bf j},\tilde {\bf j}=0},\qquad
\end{eqnarray}

\noindent
where

\begin{eqnarray}
&& (S_{PV})_I
= \sum\limits_{n=1}^\infty \frac{1}{4 n!}\int d^4x\,d^4\theta\,
\Big(\Phi^* (2V)^{n} \Phi + \tilde\Phi^* (-2V)^n\tilde\Phi\Big)
+\nonumber\\
&& \qquad\qquad\qquad\qquad
+ \Big(Z(\Lambda/\mu) - 1\Big) \frac{1}{4}\int d^4x\,d^4\theta\,
\Big(\Phi^* e^{2V} \Phi + \tilde\Phi^* e^{-2V}\tilde\Phi\Big),\qquad\quad
\end{eqnarray}

\noindent
that allows to find their perturbative expansions. Then it is possible
to calculate the generating functional $Z$ according to the prescription

\begin{eqnarray}\label{Generating_Functional_Z}
&& Z = \prod\limits_i
\Bigg(\det PV\Big(\frac{1}{i}\frac{\delta}{\delta J},M_i\Big)\Bigg)^{c_i}
\exp\Bigg\{i S_I\Bigg(\frac{1}{i}\frac{\delta}{\delta J},\,
\frac{1}{i}\frac{\delta}{\delta j},\ldots\Bigg)\Bigg\}
\times\nonumber\\
&& \qquad\qquad\qquad
\times
\exp\Bigg\{ i \int d^4x\,d^4\theta\,\Bigg(
j^* \frac{1}{\partial^2}  j
+ \tilde j^* \frac{1}{\partial^2} \tilde j
- J \frac{e^2}{\partial^2
\Big(1+\partial^{2n}/\Lambda^{2n}\Big)} J\Bigg)\Bigg\}.\qquad
\end{eqnarray}

The generating functional for connected Green functions in our notations
is written as

\begin{equation}\label{W}
W = - i\ln Z,
\end{equation}

\noindent
and the effective action is defined by making a Legender transformation

\begin{equation}\label{Gamma}
\Gamma = W - \int d^4x\,d^4\theta\,J V
- \int d^4x\,d^2\theta\, \Big(j\,\phi + \tilde j\,\tilde\phi \Big)
- \int d^4x\,d^2\bar\theta\,
\Big(j^*\phi^* + \tilde j^* \tilde\phi^* \Big),
\end{equation}

\noindent
where $J$, $j$ and $\tilde j$ should be expressed in terms of
$V$, $\phi$ and $\tilde\phi$ through solving of the equations

\begin{equation}
V = \frac{\delta W}{\delta J};\qquad
\phi = \frac{\delta W}{\delta j};\qquad
\tilde\phi = \frac{\delta W}{\delta\tilde j}.
\end{equation}

Expressions for Feinman diagrams in the coordinate representation can
be found expanding generating functional (\ref{Generating_Functional_Z})
and substituting the result into equations (\ref{W}) and (\ref{Gamma}).
Certainly, after this procedure $\Gamma$ will contain only 1PI-diagrams.
Expressions for Feinman diagrams in the momentum space can be then
obtained by Fourier transformations. Performing the calculations we
used this algorithm and tried to avoid direct application of Feinman rules
in order to be complitely sure in the correctness of numerical factors
for all diagrams. However, for the sake of completeness we formulate
Feinman rules for the considered theory, which allow to verify structure
of expressions for the diagrams.

1. External lines correspond to a factor

\begin{equation}
\prod\limits_{E}
\int \frac{d^4p_{{}_{E_V}}}{(2\pi)^4} V(p_{{}_{E_V}})
\int \frac{d^4p_{{}_{E_\phi}}}{(2\pi)^4} \phi(p_{{}_{E_\phi}})
\cdot \ldots\cdot
(2\pi)^4 \delta\Big(\sum\limits_{E} p_{{}_E}\Big),
\end{equation}

\noindent
where the index $E$ numerates external momentums.

2. Each internal line of $V$-superfield corresponds to

\begin{equation}
\frac{2e^2}{(k^2+i0) \Big(1+(-1)^n k^{2n}/\Lambda^{2n}\Big)} \,
\delta^4(\theta_1-\theta_2).
\end{equation}

3. Each internal line $\phi-\phi^*$ or $\tilde\phi-\tilde\phi^*$
corresponds to

\begin{equation}
-\frac{1}{4 (k^2+i0)}\bar D^2 D^2 \delta^4(\theta_1-\theta_2).
\end{equation}

\noindent
(Note, that in the considered theory the action is quadratic in matter
superfields, that allows to formulate Feinman rules in a bit different
manner, than for, say, Wess-Zumino model.)

4. Pauli-Villars fields are present only in the closed loops.
Each internal line $\Phi-\Phi^*$ or $\tilde\Phi-\tilde\Phi^*$
corresponds to

\begin{equation}
- \frac{1}{4(k^2-M_i^2+i0)}\, \bar D^2 D^2 \delta^4(\theta_1-\theta_2).
\end{equation}

\noindent
Internal lines $\Phi-\tilde\Phi$ and $\Phi^*-\tilde\Phi^*$ corresponds
to

\begin{equation}
\frac{M_i}{k^2-M_i^2+i0}\,\bar D^2 \delta^4(\theta_1-\theta_2)
\quad\mbox{and}\quad
\frac{M_i}{k^2-M_i^2+i0}\, D^2 \delta^4(\theta_1-\theta_2)
\end{equation}

\noindent
respectively. Also it is necessary to add
${\displaystyle - \sum\limits_i c_i}$ for each closed loop of
Pauli-Villars fields.

5. Each loop gives integration over a loop momentum
${\displaystyle \int \frac{d^4k}{(2\pi)^4}}$.

6. Each vertex gives integration over the corresponding $\theta$:
${\displaystyle \int d^4\theta}$.

7. There are numerical factors, which can be calculated expanding
generating functional (\ref{Generating_Functional_Z}).

%%%%%%%%%%%%%%%%%%%%%%%%%%%%%%%%%%%%%%%%%%%%%%%%%%%%%%%%%%%%%%%%%%%%%%%%%%%

\section{Calculation of two-loop renormgroup functions.}
\hspace{\parindent}
\label{Section_Two_Loop}

\begin{figure}[p]
\hspace*{1.5cm}
\epsfxsize12.0truecm\epsfbox{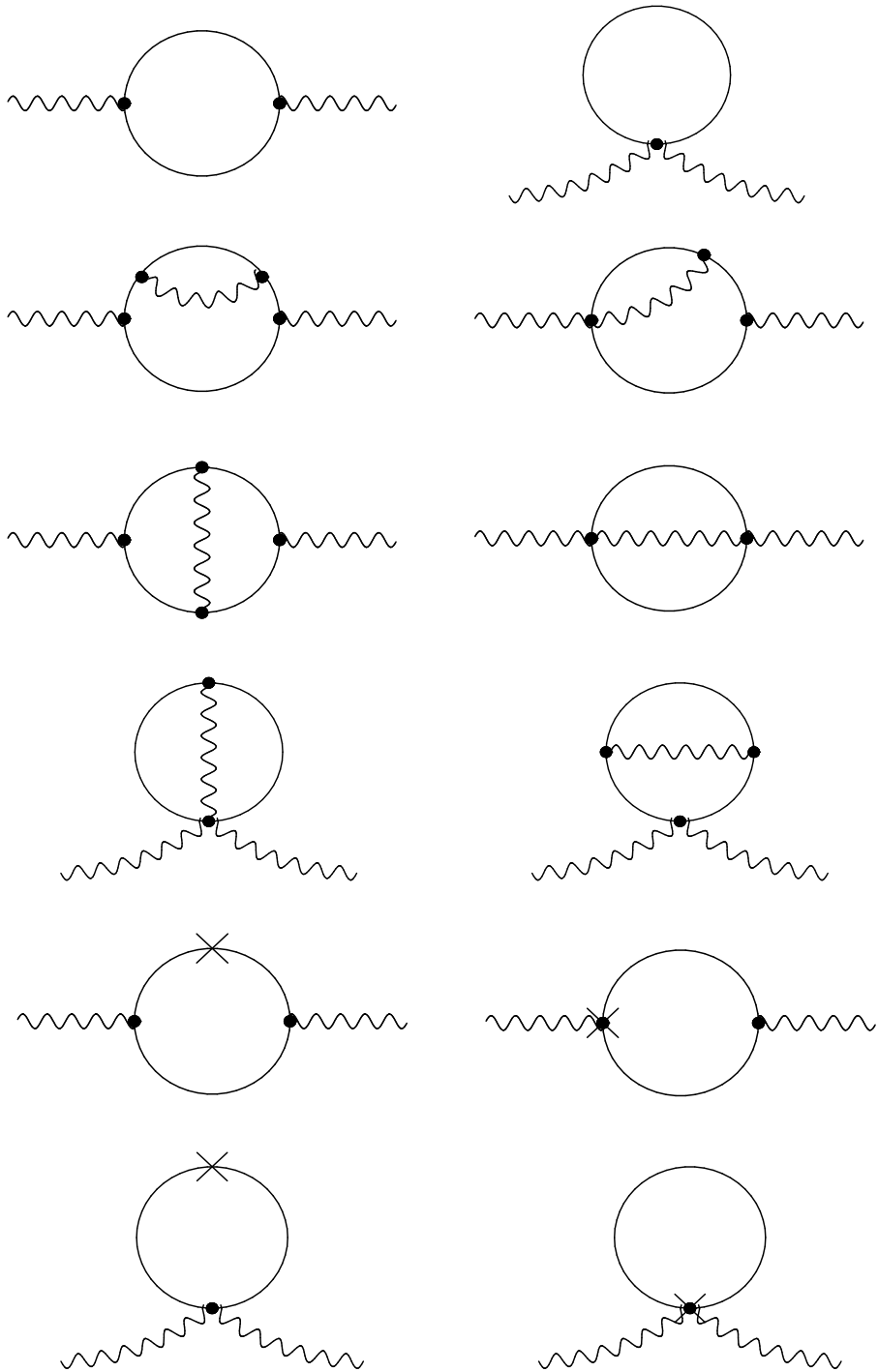}
\caption{Feinman graphs, giving nontrivial contributions to the two-loop
$\beta$-function of $N=1$ supersymmetric electrodynamics.}
\label{Figure_Beta_Diagrams}
\end{figure}

Let us calculate two-loop $\beta$-function and anomalous dimension for
a model, described by action (\ref{Regularized_SQED_Action}). In the
two-loop approximation $\beta$-function can be found after calculation
of diagrams with $E_V=2$, $E_\phi=0$, presented at Figure
\ref{Figure_Beta_Diagrams}. Note, that each graph at this figure
corresponds to a diagram with internal $\phi$-line, a diagram with
internal $\tilde\phi$-line and a set of diagrams with internal lines
of Pauli-Villars fields. As an example in Appendix \ref{Appendix_One_Loop}
we present detailed calculation of one-loop diagrams. Expressions obtained
for the other diagrams are collected in Appendix \ref{Appendix_Diagrams}.
Each of these diagrams has the following structure:

\begin{equation}
\int d^4\theta \frac{d^4p}{(2\pi)^4}
\Bigg(V(-p,\theta)\, \partial^2 \Pi_{1/2} V(p,\theta)\,f_1(p,\Lambda)
+ V(-p,\theta) V(p,\theta)\,f_2(p,\Lambda) \Bigg).
\end{equation}

\noindent
Terms proportional to
${\displaystyle \int d^4\theta\,V(-p,\theta) V(p,\theta)}$ are not
gauge invariant and should disappear after summing of all Feinman
diagrams, that is very convenient for checking correctness of the
calculations. The other terms can be written as

\begin{equation}
- \mbox{Re} \int d^2\theta \frac{d^4p}{(2\pi)^4}
W_a(-p,\theta) C^{ab} W_b(p,\theta)\,
\sum\limits_{\mbox{\footnotesize diagrams}} f_1(p,\Lambda).
\end{equation}

\noindent
Having performed the calculations, in the Minkowsky space we obtained,
that the result for two-loop contribution to the effective action,
corresponding to the two-point Green function of the gauge field, can
be written as

\begin{equation}\label{Two_Loop_Effective_Action_V}
\quad\Delta\Gamma^{(2)}_{V} = \mbox{Re} \int d^2\theta\,
\frac{d^4p}{(2\pi)^4} W_a(p) C^{ab} W_b(-p)
\Big(f_{\mbox{\footnotesize 1-loop}} + f_{\mbox{\footnotesize 2-loop}}
+ f_{\mbox{\footnotesize PV}} + f_{\mbox{\footnotesize Konishi}} \Big),
\quad
\end{equation}

\noindent
where (for simplicity we omit $+i0$ in propagators)

\begin{equation}\label{1-Loop}
f_{\mbox{\footnotesize 1-loop}} =
- \frac{i}{2}\Bigg(\int\frac{d^4k}{(2\pi)^4} \frac{1}{k^2 (k+p)^2}
- \sum\limits_i c_i \int\frac{d^4k}{(2\pi)^4}
\frac{1}{(k^2-M_i^2)\Big((k+p)^2-M_i^2\Big)}\Bigg)
\end{equation}

\noindent
is a total one-loop contribution, including contributions of diagrams
with internal lines of Pauli-Villars fields;

\begin{equation}\label{2-Loop}
f_{\mbox{\footnotesize 2-loop}}
= - e^2 \int \frac{d^4k}{(2\pi)^4}\frac{d^4q}{(2\pi)^4}
\frac{(k+p+q)^2+q^2-k^2-p^2}{k^2\Big(1 + (-1)^n k^{2n}/\Lambda^{2n}\Big)
(k+q)^2 (k+p+q)^2 q^2 (q+p)^2}
\end{equation}

\noindent
is a sum of diagrams (\ref{Two_Loop_Diagram1}) -- (\ref{Two_Loop_Diagram6})
with internal lines of $\phi$ and $\tilde\phi$ fields;

\begin{eqnarray}\label{PV}
&& f_{\mbox{\footnotesize PV}} = e^2 \sum\limits_i c_i\,
\int \frac{d^4k}{(2\pi)^4}\frac{d^4q}{(2\pi)^4}\,
\frac{1}{k^2 \Big(1 + (-1)^n k^{2n}/\Lambda^{2n}\Big)}
\times\nonumber\\
&& \times
\Bigg[\frac{(k+p+q)^2+q^2-k^2-p^2}{
\Big((k+q)^2-M_i^2\Big) \Big((k+p+q)^2-M_i^2\Big) \Big(q^2-M_i^2\Big)
\Big((q+p)^2-M_i^2\Big)}
+\nonumber\\
&& \qquad\qquad\qquad\qquad\qquad\qquad
+ \frac{4 M_i^2}{\Big((k+q)^2-M_i^2\Big) \Big(q^2-M_i^2\Big)^2
\Big((q+p)^2-M_i^2\Big)} \Bigg] \qquad\quad
\end{eqnarray}

\noindent
is a contribution of diagrams (\ref{Two_Loop_Diagram1}) --
(\ref{Two_Loop_Diagram6}) with internal lines of Pauli-Villars fields;

\begin{equation}\label{Konishi}
f_{\mbox{\footnotesize Konishi}} =
- \frac{i e^2}{2\pi^2}\ln\frac{\Lambda}{\mu}\,
\sum\limits_i c_i
\int \frac{d^4k}{(2\pi)^4}\,\frac{M_i^2}{(k^2-M_i^2)^2
\Big((k+p)^2-M_i^2\Big)}\qquad\quad
\end{equation}

\noindent
is a total contribution of diagrams (\ref{Counterterms_Diagram1}) --
(\ref{Counterterms_Diagram4}) with insersions of one-loop counterterms.
($M_i = a_i\Lambda$ are masses of Pauli-Villars fields.)

Similarly, anomalous dimension can be found after calculation
of diagrams with $E_V=0$, $E_\phi=2$, presented at Figure
\ref{Figure_Anomalous_Dimension_Diagrams}. Calculation of the one-loop
diagram is described in Appendix \ref{Appendix_One_Loop}. Results for
the other diagrams are presented in Appendix \ref{Appendix_Diagrams}.
The total two-loop contribution in the Minkowsky space can be written
in the following form:

\begin{eqnarray}\label{Two_Loop_Effective_Action_Phi}
&& \hspace*{-5mm}
\Delta\Gamma^{(2)}_{\phi} =
\int d^4\theta\,\frac{d^4p}{(2\pi)^4}\,
\Big(\phi^*(p,\theta)\,\phi(-p,\theta)
+ \tilde\phi^*(p,\theta)\,\tilde\phi(-p,\theta)\Big)
\times\nonumber\\
&& \hspace*{-5mm}
\times
\Bigg\{
i \int\frac{d^4k}{(2\pi)^4}
\frac{e^2}{2 k^2 (k+p)^2 \Big(1 + (-1)^n k^{2n}/\Lambda^{2n}\Big)}
-\nonumber\\
&& \hspace*{-5mm}
- \int \frac{d^4k}{(2\pi)^4}\,\frac{d^4q}{(2\pi)^4}\,
\frac{e^4}{k^2 q^2 (k+p)^2 (q+p)^2
\Big(1+(-1)^n k^{2n}/\Lambda^{2n}\Big)
\Big(1+ (-1)^n q^{2n}/\Lambda^{2n}\Big)}
-\nonumber\\
&&\hspace*{-5mm}
- \int \frac{d^4k}{(2\pi)^4}\,\frac{d^4q}{(2\pi)^4}\,
\frac{e^4}{\displaystyle k^2 q^2 (q+p)^2 (k+q+p)^2
\Big(1+(-1)^n k^{2n}/\Lambda^{2n}\Big)
\Big(1+(-1)^n q^{2n}/\Lambda^{2n}\Big)}
+\nonumber\\
&&\hspace*{-5mm}
+ \frac{i e^4}{4\pi^2}\ln \frac{\Lambda}{\mu}
\int \frac{d^4k}{(2\pi)^4}\,\frac{1}{k^2 (k+p)^2
\Big(1+(-1)^n k^{2n}/\Lambda^{2n}\Big)}
+ \int \frac{d^4k}{(2\pi)^4}\,\frac{d^4q}{(2\pi)^4}\,
\times\nonumber\\
&&\hspace*{-5mm}
\times
\frac{e^4 (k+q+2p)^2}{k^2 (k+p)^2 q^2 (q+p)^2 (k+q+p)^2
\Big(1+(-1)^n k^{2n}/\Lambda^{2n}\Big)
\Big(1+(-1)^n q^{2n}/\Lambda^{2n}\Big)}
-\nonumber\\
&&\hspace*{-5mm}
- \int \frac{d^4q}{(2\pi)^4}
\,\frac{e^4}{q^2 (q+p)^2 \Big(1+ (-1)^n q^{2n}/\Lambda^{2n}\Big)^2}
\Bigg(\int \frac{d^4k}{(2\pi)^4}\, \frac{1}{k^2 (k+q)^2}
-\nonumber\\
&&\qquad\qquad\qquad\qquad\qquad\qquad
- \sum\limits_i c_i
\int \frac{d^4k}{(2\pi)^4}\,\frac{1}{(k^2-M_i^2)\Big((k+q)^2-M_i^2\Big)}
\Bigg).\qquad
\end{eqnarray}

Results (\ref{Two_Loop_Effective_Action_V}) and
(\ref{Two_Loop_Effective_Action_Phi}) are evidently invariant under
supersymmetry transformations, because they can be written as
integrals from products of superfields (\ref{Phi_Superfield})
and (\ref{W_Superfield}) over the superspace. The gauge
invariance is absent in equation (\ref{Two_Loop_Effective_Action_Phi})
because the calculations were made only for diagrams with $E_V=0$,
$E_\phi=2$. Adding of terms, corresponding to diagrams with arbitrary
$E_V$ and $E_\phi=2$, will certainly restore the gauge invariance.

Divergent parts of the integrals in equations (\ref{1-Loop}) --
(\ref{Konishi}) are calculated in Appendix \ref{Appendix_Integrals}.
Using results, obtained there, one can conclude, that the sum of
contributions (\ref{1-Loop}) and (\ref{2-Loop}) gives NSVZ result
for the $\beta$-function. Expression (\ref{PV}) is shown in Appendix
\ref{Appendix_Integrals} to be a finite constant and does not contribute
to the $\beta$-function. However, sum of diagrams with insersions of
one-loop counterterms (\ref{Konishi}) is not zero and exactly cancels
contribution (\ref{2-Loop}). Actually, equation (\ref{Konishi}) produces
Konishi anomaly \cite{Konishi}, calculated by using Pauli-Villars
regularization according to a method, described in \cite{Bertlmann}.
According to this method, an anomaly is equal to a contribution of
diagrams with internal lines of Pauli-Villars fields in the limit
$M_i\to 0$, while contributions of diagrams with internal lines of usual
fields are equal to 0.

Divergent parts of integrals in equation
(\ref{Two_Loop_Effective_Action_Phi}) are also calculated in Appendix
\ref{Appendix_Integrals}.

Using results of Appendix \ref{Appendix_Integrals} it is easy to verify,
that counterterms, needed to cancel two-loop divergences in the minimal
subtraction scheme can be written as

\begin{eqnarray}
&& \Delta S = - \frac{1}{16\pi^2} \ln\frac{\Lambda}{\mu}
\,\mbox{Re}\int d^4x\,d^2\theta\,\,W_a C^{ab}
\Big(1+ \frac{\partial^{2n}}{\Lambda^{2n}}\Big) W_b
+\nonumber\\
&& + \frac{1}{4} \int d^4x\,d^4\theta\,
\Big(\phi^* e^{2V} \phi + \tilde\phi^* e^{-2V} \tilde\phi\Big)
\times\nonumber\\
&& \qquad\qquad\qquad\qquad\quad
\times \smash{\Bigg\{
\frac{\alpha}{\pi} \ln\frac{\Lambda}{\mu}
+ \frac{\alpha^2}{\pi^2} \ln^2 \frac{\Lambda}{\mu}
- \frac{\alpha^2}{\pi^2}
\ln \frac{\Lambda}{\mu}\Bigg(\sum\limits_i c_i \ln \frac{M_i}{\Lambda}
+ \frac{3}{2} \Bigg)\Bigg\} },\qquad\vphantom{\int}
\end{eqnarray}

\noindent
that corresponds to

\begin{eqnarray}\label{Two_Loop_E2}
&& \frac{4\pi^2}{e_0^2} =
\frac{\pi}{\alpha\Big(\Lambda/\mu\Big)}
- \ln\frac{\Lambda}{\mu}+O(\alpha^2);\\
\label{Two_Loop_Z}
&& Z\Big(\Lambda/\mu\Big)
= 1 + \frac{\alpha}{\pi}\,\ln\frac{\Lambda}{\mu}
+ \frac{\alpha^2}{\pi^2}\,\ln^2 \frac{\Lambda}{\mu}
- \frac{\alpha^2}{\pi^2}\,\ln\frac{\Lambda}{\mu}
\Bigg(\sum\limits_i c_i \ln \frac{M_i}{\Lambda}
+ \frac{3}{2} \Bigg)+ O(\alpha^3).\qquad
\end{eqnarray}

\noindent
Therefore the two-loop $\beta$-function and anomalous dimension of
$N=1$ supersymmetric quantum electrodynamics, regularized by higher
derivatives, are written as

\begin{eqnarray}
&& \beta = \frac{\alpha^2}{\pi} + O(\alpha^4);\nonumber\\
&& \gamma(\alpha) = - \frac{\alpha}{\pi} + \frac{\alpha^2}{\pi^2}
\Bigg(\sum\limits_i c_i \ln \frac{M_i}{\Lambda}
+ \frac{3}{2} \Bigg) + O(\alpha^3).
\end{eqnarray}

\noindent
In particular, two-loop contribution to the $\beta$-function appears
to be 0, so that the beta function is completely defined by the one-loop
approximation.

The anomalous dimension $\gamma(\alpha)$ in the two-loop approximation
does not depend on $n$ or, by other words, on a form of regularizing term.
However, it depends on the ratios of Pauli-Villars masses to the constant
$\Lambda$. Nevertheless, the dependence on $M_i/\Lambda$ can be removed by
addition of finite counterterms, proportional to $\ln M_i/\Lambda$:

\begin{eqnarray}\label{Two_Loop_S_Ren_Modefied}
&& S_{\mbox{\footnotesize ren}} =
\frac{1}{4 e^2} Z_3(\Lambda/\mu)\,\mbox{Re}\int d^4x\,d^2\theta\,W_a C^{ab}
\Big(1+ \frac{\partial^{2n}}{\Lambda^{2n}}\Big) W_b
- \frac{1}{16\pi^2}\sum\limits_i c_i \ln \frac{M_i}{\Lambda}
\times\qquad\nonumber\\
&& \times
\,\mbox{Re}\int d^4x\,d^2\theta\,W_a C^{ab} W_b
+ Z(\Lambda/\mu)\frac{1}{4}\int d^4x\, d^4\theta\,
\Big(\phi^* e^{2V}\phi +\tilde\phi^* e^{-2V}\tilde\phi\Big),\qquad
\end{eqnarray}

\noindent
that corresponds to another subtraction scheme, in which anomalous
dimension is equal to

\begin{equation}
\gamma(\alpha) = - \frac{\alpha}{\pi} + \frac{3\alpha^2}{2\pi^2}
+ O(\alpha^3)
\end{equation}

\noindent
and does not depend on both $n$ and $M_i/\Lambda$. In principle, in
this scheme it is possible to consider, that $\Lambda\to\infty$,
$M_i/\Lambda\to\infty$ instead of $M_i = a_i\Lambda$.

%%%%%%%%%%%%%%%%%%%%%%%%%%%%%%%%%%%%%%%%%%%%%%%%%%%%%%%%%%%%%%%%%%%%%%%%%%

\section{Comparing of the results with predictions of the renormalization
group method.}
\label{Section_Renormgroup}
\hspace{\parindent}

The obtained results can be checked by the renormalization group method.
It is well known \cite{Collins}, that in renormalizable theories terms
proportional to $\ln^2\mu/\Lambda$ are completely defined by one-loop
counterterms. Therefore, it is possible to calculate such terms by
renormgroup equations and compare them with the result of calculation of
Feinman graphs.

Using the notation

\begin{equation}
t = \ln \frac{\mu}{\Lambda}
\end{equation}

\noindent
for the considered model renormgroup equations can be written as

\begin{equation}\label{Renormgroup}
Z(t) = \exp\Big\{ \int dt\,\gamma\Big(\alpha(t)\Big) \Big\};\qquad
t = \int \frac{d\alpha}{\beta(\alpha)}.
\end{equation}

\noindent
Because in the one-loop approximation the $\beta$-function is equal to

\begin{equation}
\beta(\alpha) = \alpha^2 \beta_1 + O(\alpha^3),
\end{equation}

\noindent
in the lowest order

\begin{equation}
\alpha(t) = \alpha_0 \Big(1+ \beta_1 \alpha_0\, t + O(\alpha_0^2)\Big),
\end{equation}

\noindent
where $\alpha_0 = \alpha(0)$. Expanding the anomalous dimension in powers
of $\alpha$

\begin{equation}
\gamma(\alpha) = \alpha \gamma_1 + \alpha^2 \gamma_2 + O(\alpha^3)
= \gamma_1\Big(\alpha_0 + \beta_1 \alpha_0^2\, t\Big)
+ \alpha_0^2\,\gamma_2 + O(\alpha_0^3)
\end{equation}

\noindent
and substituting it to the first equation of (\ref{Renormgroup}),
we obtain, that

\begin{equation}
Z(t) = 1 + \gamma_1 \alpha_0\,t + \gamma_1 \beta_1\alpha_0^2\,t^2/2
+ \gamma_2 \alpha_0^2\,t + \gamma_1^2 \alpha_0^2\,t^2/2 + O(\alpha_0^3).
\end{equation}

\noindent
Taking into account, that according to the results of one-loop
calculations $\gamma_1 = - 1/\pi$ and $\beta_1 = 1/\pi$, the function
$Z$ should take the following form:

\begin{eqnarray}
&& Z(\Lambda/\mu) = 1 + \frac{\alpha_0}{\pi}\, \ln \frac{\Lambda}{\mu}
- \gamma_2 \alpha_0^2\, \ln \frac{\Lambda}{\mu} + O(\alpha_0^3)
=\nonumber\\
&& \qquad\qquad\qquad\qquad
= 1 + \frac{\alpha}{\pi}\,\ln \frac{\Lambda}{\mu}
+ \frac{\alpha^2}{\pi^2}\,\ln^2 \frac{\Lambda}{\mu}
- \gamma_2 \alpha^2\, \ln \frac{\Lambda}{\mu} + O(\alpha^3).\qquad
\end{eqnarray}

\noindent
Comparing this expression with equation (\ref{Two_Loop_Z}) we see, that
terms proportional to $\ln^2\mu/\Lambda$ coincide, that can be considered
as a check of performed calculations.

%%%%%%%%%%%%%%%%%%%%%%%%%%%%%%%%%%%%%%%%%%%%%%%%%%%%%%%%%%%%%%%%%%%%%%%%%%

\section{Conclusion.}
\hspace{\parindent}

In this paper we calculated two-loop $\beta$-function and anomalous
dimension for $N=1$ supersymmetric massless quantum electrodynamics,
regularized by higher derivatives. In particular, two-loop contribution
to the $\beta$-function is found to be 0 and not to depend on the form of
higher derivative term. As we mentioned above, this result follows from
the fact, that the axial anomaly and the anomaly of energy-momentum
tensor trace in the considered model belong to one supermultiplet.
However, to obtain it we have to perform calculations using higher
covariant derivative regularization. In principle, this regularization
(complemented by the Pauli-Villars regularization for one-loop diagrams)
allows to perform easy calculation of diagrams with insertion of
counterterms, which are proportional to Konishi anomaly \cite{Konishi}
and have nonzero contribution. Possibly the results of the paper allow
to assume, that contributions of all higher loops to the $\beta$-function
of $N=1$ supersymmetric electrodynamics, regularized by higher derivatives,
are also equal to 0. However, to be completely sure in it, it is necessary
to calculate scheme dependent three-loop $\beta$-function.

Two-loop anomalous dimension is found to be independent on the form of
higher derivative term if one renormalizes the coupling constant in this
term. However, $\gamma(\alpha)$ depends on the ratios of Pauli-Villars
masses to the constant $\Lambda$. Nevertheless, it is possible to make
anomalous dimension completely independent on parameters of higher
derivative regularization ($n$ and $M/\Lambda$) if one introduces some
finite counterterms in the renormalized action, that actually corresponds
to a different choice of renormalization scheme.

Note, that the result for $\beta$-function does not contradict to
NSVZ result

\begin{equation}
\beta(\alpha) = \frac{\alpha^2}{\pi^2}\Big(1-\gamma(\alpha)\Big),
\end{equation}

\noindent
because in considered model $\beta$-function depends
on the normalization of the matter superfields \cite{Arkani}. In
particular, after a scale transformation making matter superfields
canonically normalized, it is possible to obtain NVSZ $\beta$-function.
However, in the present paper this statement was not checked by
explicit calculations and we hope to make it later.

%%%%%%%%%%%%%%%%%%%%%%%%%%%%%%%%%%%%%%%%%%%%%%%%%%%%%%%%%%%%%%%%%%%%%%%%%%

\vspace{1cm}

\noindent
{\Large{\bf Acknowledgments}}

\bigskip

\noindent
Authors are very grateful to P.I.Pronin and A.A.Slavnov for valuable
discussions.

\vspace{1cm}

\noindent
{\Large\bf Appendix.}

%\pagebreak

\appendix

%%%%%%%%%%%%%%%%%%%%%%%%%%%%%%%%%%%%%%%%%%%%%%%%%%%%%%%%%%%%%%%%%%%%%%%%%%%

\section{Superficial degree of divergence for the supersymmetric
quantum electrodynamics, regularized by higher derivatives.}
\hspace{\parindent}
\label{Appendix_Degree_Of_Divergence}

In order to calculate the superficial degree of divergence for an
arbitrary diagram in the massless supersymmetric quantum electrodynamics
let us introduce the following notations:

$L$ is a number of loops,

%$N_V$ is a number of vertexes,

$I_V$ is a number of internal $V$-lines,

$I_\phi$ is a number of internal $\phi$-lines,

$E_V$ is a number of external $V$-lines,

$E_\phi$ is a number of external $\phi$-lines.

%According to the Feinman rules, described above, each scalar propagator
%behaves as $k^{-2}$, while each propagator of $V$-superfield behaves as
%$k^{-2-2n}$ due to existance of a term with higher derivatives. Each
%vertex without external scalar lines contains two factors $D^2$ and each
%vertex with an external scalar line contains one factor $D^2$. Because
%dimension of $D^2$ is equal to dimension of $k$, such vertexes increase
%degree of divergence by 2 or 1 respectively.

%Each loop contains integration over a loop momentum ($k^{4L}$). However,
%it is necessary to use 2 factors $D^2$ in order to eliminate products of
%$\delta$-functions for each loop.

%Therefore, taking into account identities
%(\ref{Appendix_Degree_Identities}), the superficial degree of divergence
%can be written as

%\begin{eqnarray}\label{Appendix_Degree_Degree_Of_Divergence}
%&& \omega_\Lambda = 2 L - 2 I_\phi - I_V (2 + 2n) + 2 N_V - E_\phi
%=\nonumber\\
%&& \qquad\qquad\qquad\qquad
%= 2 - 2n (L-1) - E_\phi (n+1).\qquad
%\end{eqnarray}

We will start with the result for the superficial degree of divergence
of massless supersymmetric electrodynamics without higher derivatives
\cite{Ferrara2}

\begin{equation}
\omega = 2 - E_\phi.
\end{equation}

\noindent
Adding of the higher derivative term changes only the propagator of
$V$-superfield, which will be proportional to $k^{-2-2n}$ instead of
$k^{-2}$ in the usual supersymmetric electrodynamics. Therefore, in the
regularized theory the superficial degree of divergence is equal to

\begin{equation}\label{Degree_Divergence}
\omega_\Lambda = \omega - 2 n I_V.
\end{equation}

Taking into account, that $\phi$-lines are continuous, it is easy
to prove the following identity:

\begin{eqnarray}\label{Appendix_Degree_Identities}
%&& N_V = I_\phi + \frac{1}{2} E_\phi;\nonumber\\
&& L = I_V + 1 - \frac{1}{2} E_\phi.
\end{eqnarray}

\noindent
Therefore expression (\ref{Degree_Divergence}) can be finally rewritten
as

\begin{equation}\label{Appendix_Degree_Degree_Of_Divergence}
\omega_\Lambda = 2 - 2n (L-1) - E_\phi (n+1).\qquad
\end{equation}

\noindent
In principle this result can be also obtaind from the Feinman rules in
the superspace, but the derivation is more complicated.

%%%%%%%%%%%%%%%%%%%%%%%%%%%%%%%%%%%%%%%%%%%%%%%%%%%%%%%%%%%%%%%%%%%%%%%%%%%

\section{Calculation of one-loop $\beta$-function and anomalous
dimension.}
\hspace{\parindent}
\label{Appendix_One_Loop}

Expanding generating functional (\ref{Generating_Functional_Z}) and
substituting it to effective action (\ref{Gamma}) it is possible to
find, that a one-loop diagram with $E_V=0$ and $E_\phi=2$ in the
coordinate representation is written as

\begin{eqnarray}\label{One_Loop_Phi_Diagram}
&& \smash{\epsfxsize6.0truecm\epsfbox[190 390 490 890]{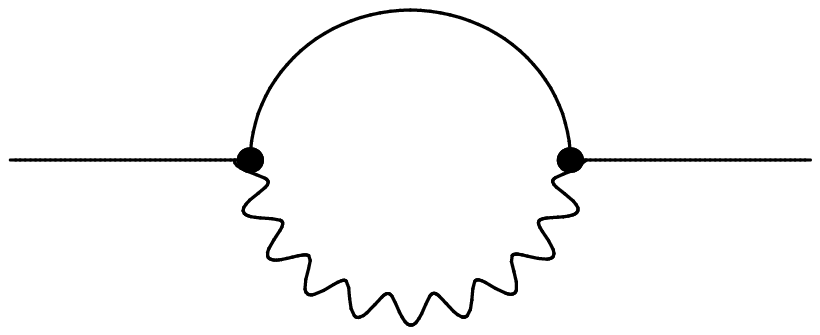}}
\hspace*{-1.2cm}
=\vphantom{\int\limits_p}\nonumber\\
&& \vphantom{\int\limits^h}
= \frac{i}{8}\int d^8x_1 d^8x_2\,
\phi^*(x_1,\theta_1)\,\phi(x_2,\theta_2)
\frac{e^2}{\partial^2 \Big(1+\partial^{2n}/\Lambda^{2n}\Big)}
\delta^8_{12}
\frac{D^2 \bar D^2}{\partial^2} \delta^8_{12}.\qquad
\end{eqnarray}

\noindent
where

\begin{equation}
\int d^8x \equiv \int d^4x\, d^4\theta;\qquad
\delta^8_{12} \equiv \delta^4(x_1-x_2)\,\delta^4(\theta_1-\theta_2).
\end{equation}

\noindent
After Fourier transformation this expression in the Minkowsky
space can be written as

\begin{eqnarray}\label{One_Loop_Anomalous_Dimension_Calculation}
&& \frac{i}{8}\int d^4\theta_1 d^4\theta_2
\int \frac{d^4p}{(2\pi)^4}\, \phi^*(p,\theta_1)\,\phi(-p,\theta_2)
\int \frac{d^4k}{(2\pi)^4}
\times\nonumber\\
&&\qquad\qquad\quad
\times
\frac{e^2}{k^2 \Big(1+(-1)^n k^{2n}/\Lambda^{2n}\Big)}
\delta^4(\theta_1-\theta_2)
\frac{1}{(k+p)^2} D^2 \bar D^2 \delta^4(\theta_1-\theta_2)
=\qquad\nonumber\\
&& = \frac{i}{2}\int d^4\theta \frac{d^4p}{(2\pi)^4}
\frac{d^4k}{(2\pi)^4}\,\phi^*(p,\theta)\,\phi(-p,\theta)
\frac{e^2}{k^2 \Big(1 + (-1)^n k^{2n}/\Lambda^{2n}\Big) (k+p)^2},
\end{eqnarray}

\noindent
where we used the identity

\begin{equation}\label{Delta_Theta_Identity}
\delta^4(\theta_1-\theta_2)
\,D^2 \bar D^2 \delta^4(\theta_1-\theta_2)
=  4\,\delta^4(\theta_1-\theta_2).
\end{equation}

The integral over $d^4k$ in equation
(\ref{One_Loop_Anomalous_Dimension_Calculation}) can be calculated
after the Wick rotation:

\begin{equation}
\frac{i}{2}\int \frac{d^4k}{(2\pi)^4}\,
\frac{e^2}{k^2 \Big(1 + (-1)^n k^{2n}/\Lambda^{2n}\Big) (k+p)^2}
\to
- \frac{1}{2}\int \frac{d^4k}{(2\pi)^4}
\frac{e^2}{k^2 \Big(1 + k^{2n}/\Lambda^{2n}\Big) (k+p)^2}.
\end{equation}

\noindent
Then it is possible to perform calculation in the Eucliedian space and
analitically continue the result for imaginary $p^0$.

The diagrams contributing to the one-loop $\beta$-function can be
considered similarly.

From equations (\ref{Generating_Functional_Z}), (\ref{W}) and
(\ref{Gamma}) it is possible to find, that in the coordinate
representation

\bigskip

\begin{eqnarray}
&& \smash{\epsfxsize6.0truecm\epsfbox[240 365 540 865]{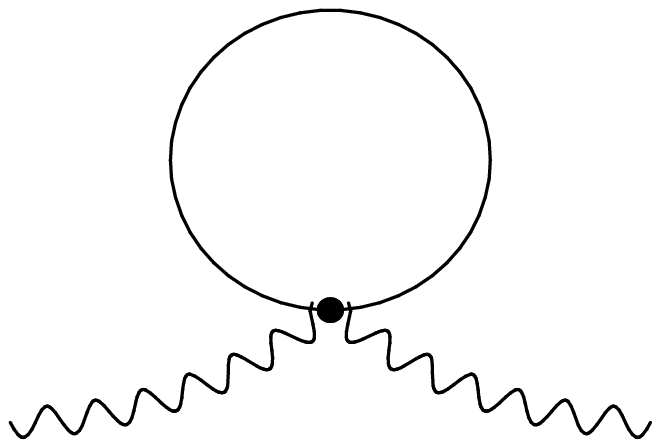}}
\hspace*{-3.4cm}
= -i\int d^8x_1 V(x_1)^2 \frac{D_1^2 \bar D_1^2}{4 \partial^2}\delta^8_{11}
= -i\int d^8x_1 V(x_1)^2 \frac{1}{\partial^2}\delta^4(x_1-x_1)
=\nonumber\\
&& \qquad\qquad\qquad\qquad\qquad\qquad\qquad
= i\int d^4\theta \frac{d^4p}{(2\pi)^4}\,V(p,\theta)\,V(-p,\theta)
\int \frac{d^4k}{(2\pi)^4} \frac{1}{k^2},\quad
\end{eqnarray}

\noindent
where we take into account identity (\ref{Delta_Theta_Identity}).
The corresponding diagram with Pauli-Villars fields in the coordinate
representation is written as

\bigskip

\begin{eqnarray}
&& \smash{\epsfxsize6.0truecm\epsfbox[240 365 540 865]{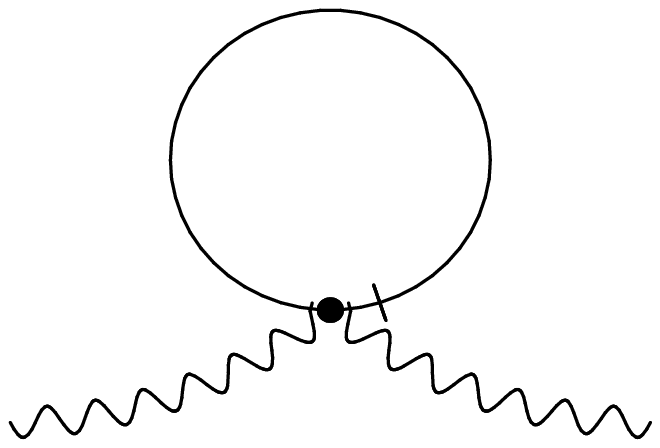}}
\hspace*{-3.4cm}
= i\sum\limits_i c_i \int d^8x_1 V(x_1)^2
\frac{D_1^2 \bar D_1^2}{4 (\partial^2+M_i^2)}\delta^8_{11}
=\\
&& \qquad\qquad\qquad\qquad\qquad
= - i\sum\limits_i c_i
\int d^4\theta \frac{d^4p}{(2\pi)^4}\,V(p,\theta)\,V(-p,\theta)
\int \frac{d^4k}{(2\pi)^4} \frac{1}{k^2-M_i^2}.\quad\nonumber
\end{eqnarray}

\noindent
(A bar at diagrams with Pauli-Villarse fields denotes a part of a vertex
corresponding to $\Phi^*$ or $\tilde\Phi^*$.)

For the next diagram equations (\ref{Generating_Functional_Z}), (\ref{W})
and (\ref{Gamma}) give the following expression:

\begin{eqnarray}\label{Beta_Diagram1}
&&\quad\ \ \smash{\epsfxsize6.0truecm\epsfbox[270 390 570 890]{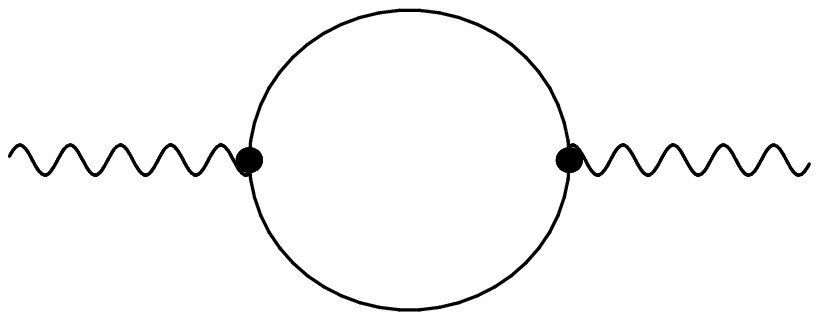}}
\hspace*{-2.8cm}
= -\frac{i}{4} \int d^8x_1 d^8x_2\,V(x_1)\,V(x_2)\,
\frac{D_1^2 \bar D_1^2}{4 \partial^2} \delta^8_{12}\,
\frac{\bar D_1^2 D_1^2}{4 \partial^2} \delta^8_{12}
=\vphantom{\int\limits_p}\nonumber\\
&& \vphantom{\int\limits^h}\hspace*{-1.2cm}
= -\frac{i}{4} \int d^8x_1 d^8x_2\,V(x_2)\,
\frac{1}{4\partial^2}\delta^8_{12}\,
\bar D_1^2 D_1^2 \Big(V(x_1)
\frac{\bar D_1^2 D_1^2}{4 \partial^2} \delta^8_{12}\Big)
=\nonumber\\
&& \hspace*{-1.2cm}
= -\frac{i}{4} \int d^8x_1 d^8x_2\,V(x_2)\,
\frac{1}{4\partial^2}\delta^8_{12}\,
\Big(D_1^2 \bar D_1^2 + [\bar D_1^2, D_1^2]\Big) \Big(V(x_1)
\frac{\bar D_1^2 D_1^2}{4 \partial^2} \delta^8_{12}\Big)
=\\
&& \hspace*{-1.2cm}
= -\frac{i}{4} \int d^8x_1 d^8x_2\,V(x_2)\,
\frac{1}{4\partial^2}\delta^8_{12}\,
\Bigg( D_1^2 \bar D_1^2 V(x_1)
\frac{\bar D_1^2 D_1^2}{4 \partial^2} \delta^8_{12}
+\nonumber\\
&& \qquad\qquad\qquad\qquad
+ \partial_\mu \Big(
i \bar D_1\gamma^\mu\gamma_5 D_1 V(x_1)
\frac{\bar D_1^2 D_1^2}{\partial^2} \delta^8_{12}
- 4 V(x_1)\, \partial_\mu
\frac{\bar D_1^2 D_1^2}{\partial^2} \delta^8_{12}
\Big)\Bigg).\nonumber
\end{eqnarray}

\noindent
Here we took into accout, that

\begin{equation}
\delta^4(\theta_1-\theta_2)\, D_{a_1} \ldots D_{a_k}
\delta^4(\theta_1-\theta_2) = 0
\end{equation}

\noindent
for $k=0,1,2,3$ and used identities

\begin{equation}
[\bar D^2, D^2] = 4i \bar D\gamma^\mu \gamma_5 D\,\partial_\mu;
\qquad
\bar D\gamma_\mu\gamma_5 D\, \bar D^2
= 4i \partial_\mu \bar D^2.
\end{equation}

Taking into accont equation (\ref{Delta_Theta_Identity}) in the momentum
representation expression (\ref{Beta_Diagram1}) can be written as

\begin{eqnarray}
&& - \frac{i}{16}\int d^4\theta \frac{d^4p}{(2\pi)^4}\frac{d^4k}{(2\pi)^4}
\frac{1}{k^2 (k+p)^2} V(p,\theta)
\times\nonumber\\
&& \qquad\qquad\qquad\qquad
\times
\Big(D^2 \bar D^2 + 4\bar D\gamma^\mu\gamma_5 D\,k_\mu
+ 16 (p+k)^\mu k_\mu\Big) V(-p,\theta).\qquad\vphantom{\frac{1}{2}}
\end{eqnarray}

\noindent
Let us note, that the integral

\begin{equation}
I_\mu \equiv \int \frac{d^4k}{(2\pi)^4}\frac{k_\mu}{k^2 (k+p)^2}
\end{equation}

\noindent
is proportional to $p_\mu$. Therefore, it can be presented as

\begin{eqnarray}
&& I_\mu = \frac{p_\mu p^\nu}{p^2} I_\nu
= \frac{p_\mu}{2 p^2}\int \frac{d^4k}{(2\pi)^4}\,
\frac{(k+p)^2-k^2-p^2}{k^2 (k+p)^2}
=\nonumber\\
&& \qquad\qquad\quad
= - \frac{1}{2} p_\mu \int \frac{d^4k}{(2\pi)^4}\frac{1}{k^2 (k+p)^2}
+ \frac{p_\mu}{2 p^2} \int \frac{d^4k}{(2\pi)^4}
\Bigg(\frac{1}{k^2}-\frac{1}{(k+p)^2}\Bigg).\qquad
\end{eqnarray}

\noindent
The last term can be omitted, because with the corresponding contribution
of the diagram with Pauli-Villars fields it is proportional to

\begin{equation}
\quad\int \frac{d^4k}{(2\pi)^4}
\Bigg(\frac{1}{k^2} - \sum\limits_i c_i\,\frac{1}{k^2-M_i^2}\Bigg)
- \int \frac{d^4k}{(2\pi)^4} \Bigg(\frac{1}{(k+p)^2}
- \sum\limits_i c_i\,\frac{1}{(k+p)^2-M_i^2}\Bigg) = 0,\quad
\end{equation}

\noindent
where we take into account, that both integrals are convergent and it
is possible to make in the second integral a substitution $k+p\to k$.
Therefore, the considered diagram takes the following form:

\begin{eqnarray}
&& -\frac{i}{16}\int \frac{d^4k}{(2\pi)^4}\frac{1}{k^2 (k+p)^2}
V(p,\theta)\Big(D_1^2 \bar D_1^2 - 2 \bar D\gamma^\mu\gamma_5 D\,p_\mu
+ 16 (k+p)^\mu k_\mu\Big) V(-p,\theta)
=\nonumber\\
&& = -\frac{i}{16}\int \frac{d^4k}{(2\pi)^4}\frac{1}{k^2 (k+p)^2}
V(p,\theta)\Big(D_1^2 \bar D_1^2 + \frac{1}{2} [\bar D_1^2,D_1^2]
+ 16 (k+p)^\mu k_\mu\Big) V(-p,\theta)
=\nonumber\\
%&& = - \frac{i}{2} \int \frac{d^4k}{(2\pi)^4}\frac{1}{k^2 (k+p)^2}
%V(p,\theta)\Big(- \partial^2 \Pi_{1/2} + p^2
%+ 2 (k+p)^\mu k_\mu\Big) V(-p,\theta)
%=\nonumber\\
&& = \frac{i}{2} \int \frac{d^4k}{(2\pi)^4}\frac{1}{k^2 (k+p)^2}
V(-p,\theta)\Big(\partial^2 \Pi_{1/2} - (k+p)^2 - k^2 \Big) V(p,\theta).
\end{eqnarray}

This diagram has two corresponding diagrams with the loop of Pauli-Villars
fields, presented below. The first diagram is calculated similar to the
massless case:

\begin{eqnarray}
&& \smash{\epsfxsize6.0truecm\epsfbox[220 390 520 890]{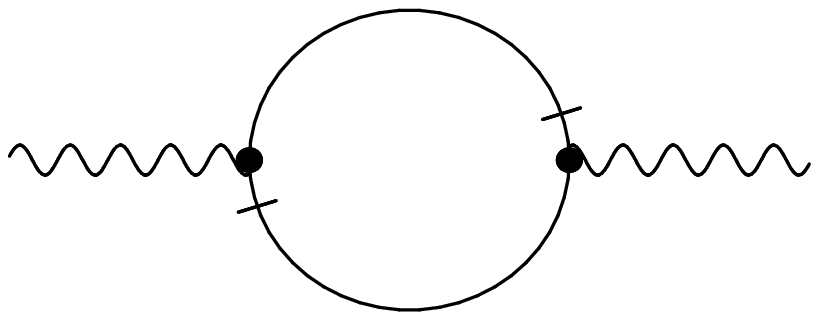}}
\hspace*{-1.8cm}
= \frac{i}{4} \sum\limits_i c_i \int d^4x_1 d^4x_2\,V(x_1)\,V(x_2)\,
\frac{D_1^2 \bar D_1^2}{4 (\partial^2+M_i^2)} \delta^8_{12}
\times\vphantom{\int\limits_p}\nonumber\\
&& \vphantom{\int\limits^h}\times
\frac{\bar D_1^2 D_1^2}{4 (\partial^2+M_i^2)} \delta^8_{12}
= \frac{i}{2} \sum\limits_i c_i \int d^4\theta \frac{d^4p}{(2\pi)^4}
\frac{d^4k}{(2\pi)^4}
\frac{1}{\Big(k^2-M_i^2\Big)\Big((k+p)^2-M_i^2\Big)}
\times\qquad\nonumber\\
&& \times
V(p,\theta) \Big(-\partial^2\Pi_{1/2} + (k+p)^2 + k^2 \Big) V(-p,\theta).
\vphantom{\Bigg)}
\end{eqnarray}

\noindent
Taking into account, that signs in $\phi^* V \phi$ and
$\tilde\phi^* V \tilde\phi$ vertexes are different, the second diagram
can be written in the following form:

\begin{eqnarray}
&& \smash{\epsfxsize6.0truecm\epsfbox[220 390 520 890]{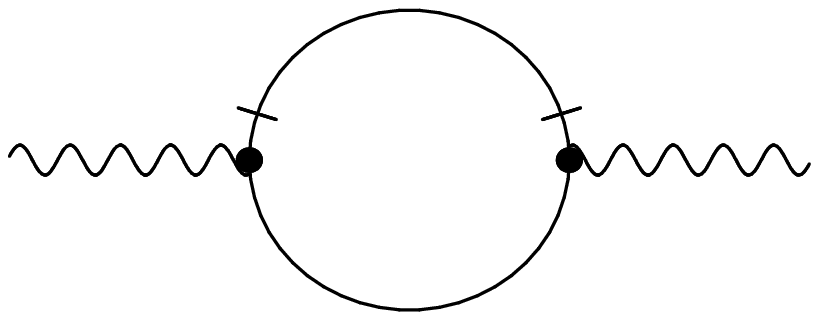}}
\hspace*{-1.8cm}
= -\frac{i}{4}\sum\limits_i c_i
\int d^8x_1 d^8x_2\,V(x_1)\,V(x_2)\,
\frac{M_i D_1^2}{\partial^2+M_i^2}\delta^8_{12}
\times\vphantom{\int\limits_p}\nonumber\\
&& \vphantom{\int\limits^h}\times
\frac{M_i \bar D_1^2}{\partial^2+M_i^2} \delta^8_{12}
= -\frac{i}{4}\sum\limits_i c_i
\int d^4\theta_1 d^4\theta_2
\frac{d^4p}{(2\pi)^4} \frac{d^4k}{(2\pi)^4}\,
V(p,\theta_1)\,V(-p,\theta_2)\,
\frac{M_i^2}{k^2-M_i^2}
\times\nonumber\\
&& \times
\delta^4(\theta_1-\theta_2)\,\frac{1}{(k+p)^2-M_i^2}\,
\bar D_1^2 D_1^2
\delta^4(\theta_1-\theta_2)
= -i \sum\limits_i c_i
\int d^4\theta \frac{d^4p}{(2\pi)^4} \frac{d^4k}{(2\pi)^4}\,
\times\nonumber\\
&& \times
V(p,\theta)\, V(-p,\theta)\,
\frac{M_i^2}{\Big(k^2-M_i^2\Big)\Big((k+p)^2-M_i^2\Big)}.
\end{eqnarray}

\noindent
Collecting the results we see, that the sum of the one-loop diagrams
is equal to

\begin{eqnarray}
&& \frac{i}{2}\int d^4\theta \frac{d^4p}{(2\pi)^2}\,
V(p,\theta)\,\partial^2\Pi_{1/2}V(-p,\theta)\,
\times\nonumber\\
&& \qquad\qquad\quad
\times
\int \frac{d^4k}{(2\pi)^2}\Bigg(\frac{1}{k^2 (k+p)^2}
-\sum\limits_i c_i
\frac{1}{\Big(k^2-M_i^2\Big)\Big((k+p)^2-M_i^2\Big)}\Bigg)
=\nonumber\\
&& = - \frac{i}{2}\mbox{Re} \int d^2\theta \frac{d^4p}{(2\pi)^2}\,
W_a(p,\theta) C^{ab} W_b(-p,\theta)\,
\times\nonumber\\
&& \qquad\qquad\quad
\times
\int \frac{d^4k}{(2\pi)^2}\Bigg(\frac{1}{k^2 (k+p)^2}
-\sum\limits_i c_i
\frac{1}{\Big(k^2-M_i^2\Big)\Big((k+p)^2-M_i^2\Big)}\Bigg).
\qquad
\end{eqnarray}

\noindent
Note, that all noninvariant terms, proportional to $V^2$, disappeared,
that can be considered as a check of the correctness of the calculations.

After Wich rotation in the Eucliedian space the last integral over $d^4k$
can be written as

\begin{equation}
\frac{1}{2}\int \frac{d^4k}{(2\pi)^2}\Bigg(\frac{1}{k^2 (k+p)^2}
-\sum\limits_i c_i
\frac{1}{\Big(k^2+M_i^2\Big)\Big((k+p)^2+M_i^2\Big)}\Bigg).
\end{equation}

\noindent
This integral is calculated below in Appendix \ref{Appendix_Integrals}.
In order to find the contribution to the effective action the result
should be continued for imaginary $p_0$ and multiplied by

\begin{equation}
\mbox{Re}\int d^2\theta \frac{d^4p}{(2\pi)^4}\,
W_a(p,\theta)C^{ab} W_b(-p,\theta).
\end{equation}

Two-loop Feinman diagrams are calculated in the similar way. However,
the calculations are much more complicated and we do not describe them
in details.

%%%%%%%%%%%%%%%%%%%%%%%%%%%%%%%%%%%%%%%%%%%%%%%%%%%%%%%%%%%%%%%%%%%%%%%%%%%

\section{Diagrams, giving nontrivial contribution to the two-loop
$\beta$-function and anomalous dimension.}
\hspace{\parindent}
\label{Appendix_Diagrams}

Below we present expressions for all Feinman graphs, giving nontrivial
contributions to the two-loop $\beta$-function. Each of these graphs
corresponds to a set of diagrams, which consists of a diagram with
internal $\phi$-line, a diagram with internal $\tilde\phi$-line and
diagrams with internal lines of Pauli-Villars fields. In order to
find contributions to the effective action it is necesary to add the
factor

\begin{equation}
\int d^4\theta \frac{d^4p}{(2\pi)^4}.
\end{equation}

\noindent
Note, that for simplicity of notations we also omit $+i0$ in propagators.

\begin{eqnarray}
\label{One_Loop_Diagram1}
&& \smash{\epsfxsize6.0truecm\epsfbox[190 390 490 890]{dg1.eps}}
\hspace*{-1.2cm} = \frac{i}{2} \int \frac{d^4k}{(2\pi)^4}\Bigg[
\, V\partial^2\Pi_{1/2} V \frac{1}{k^2 (k+p)^2}
- V^2 \frac{1}{k^2}
-\nonumber\\
&& - V^2 \frac{1}{(k+p)^2}
- \sum\limits_i c_i \Bigg(
V\partial^2\Pi_{1/2} V \frac{1}{(k^2-M_i^2)\Big((k+p)^2-M_i^2\Big)}
- V^2 \frac{1}{k^2-M_i^2}
-\nonumber\\
&& - V^2 \frac{1}{(k+p)^2-M_i^2}
\Bigg)\Bigg];
\\
\nonumber\\
\nonumber\\
\nonumber\\
\label{One_Loop_Diagram2}
&& \smash{\epsfxsize6.0truecm\epsfbox[190 365 490 865]{dg6.eps}}
\hspace*{-2.0cm}
= i \int \frac{d^4k}{(2\pi)^4} V^2
\Bigg(\frac{1}{k^2}-\sum\limits_i c_i\,\frac{1}{k^2-M_i^2}\Bigg);
\\
\nonumber\\
\nonumber\\
\nonumber\\
\label{Two_Loop_Diagram1}
&&
\smash{\epsfxsize6.0truecm\epsfbox[190 390 490 890]{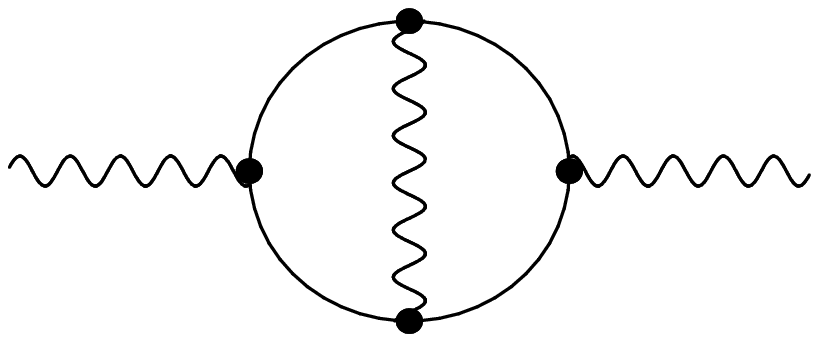}}
\hspace*{-1.2cm} =
\int \frac{d^4k}{(2\pi)^4}\frac{d^4q}{(2\pi)^2}
\,\frac{e^2}{k^2 \Big(1 + (-1)^n k^{2n}/\Lambda^{2n}\Big)}
\times\vphantom{\int\limits_p}\nonumber\\
&& \times
\Bigg[
\, V \partial^2\Pi_{1/2} V
\frac{4(k+p+q)^2-k^2-p^2}{
(k+q)^2 (k+p+q)^2 q^2 (q+p)^2}
- V^2 \frac{2}{(k+q)^2 q^2} \Bigg]
-\nonumber\\
&& -\sum\limits_i c_i
\int \frac{d^4k}{(2\pi)^4}\frac{d^4q}{(2\pi)^2}
\,\frac{e^2}{k^2 \Big(1 + (-1)^n k^{2n}/\Lambda^{2n}\Big)}
\Bigg[
\,V\partial^2\Pi_{1/2}V
\times\nonumber\\
&& \times
\frac{4(k+p+q)^2-k^2-p^2-2 M_i^2}{
\Big((k+q)^2-M_i^2\Big) \Big((k+p+q)^2-M_i^2\Big) \Big(q^2-M_i^2\Big)
\Big((q+p)^2-M_i^2\Big)}
-\nonumber\\
&&\qquad\qquad\qquad\qquad\qquad\qquad\qquad\qquad\quad\
- V^2 \frac{2}{\Big((k+q)^2-M_i^2\Big) \Big(q^2-M_i^2\Big)}
\Bigg];
\\
\nonumber\\
\nonumber\\
\label{Two_Loop_Diagram2}
&&
\smash{\epsfxsize6.0truecm\epsfbox[190 390 490 890]{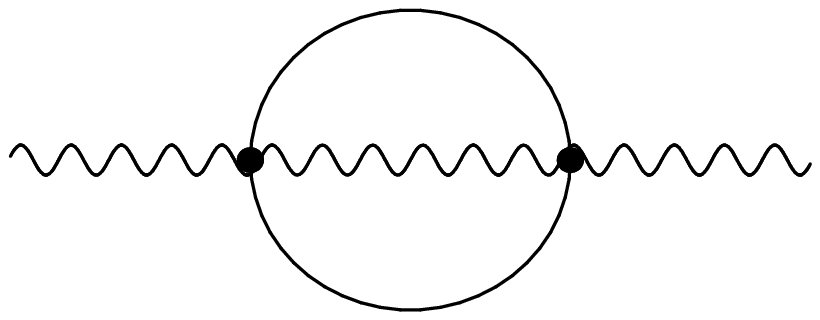}}
\hspace*{-1.2cm} = \int \frac{d^4k}{(2\pi)^4}\frac{d^4q}{(2\pi)^4}\, V^2
\frac{e^2}{k^2 \Big(1 + (-1)^n k^{2n}/\Lambda^{2n}\Big)}
\times\\
&& \qquad\qquad\qquad\quad
\times
\Bigg[- \frac{2}{(k+q)^2 (q+p)^2}
+\sum\limits_i c_i
\frac{2}{\Big((k+q)^2-M_i^2\Big) \Big((q+p)^2-M_i^2\Big)} \Bigg];
\nonumber\\
\nonumber\\
\nonumber\\
\label{Two_Loop_Diagram3}
&&
\smash{\epsfxsize6.0truecm\epsfbox[190 390 490 890]{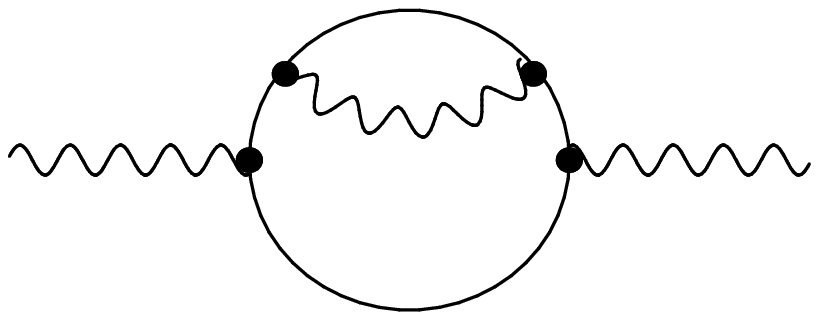}}
\hspace*{-1.2cm}
= \int \frac{d^4k}{(2\pi)^4}\frac{d^4q}{(2\pi)^4}
\frac{e^2}{k^2 \Big(1+(-1)^n k^{2n}/\Lambda^{2n}\Big)}
\times\nonumber\\
&& \times
\Bigg[
\,V\partial^2\Pi_{1/2}V \frac{2}{(k+q)^2 q^2 (q+p)^2}
- V^2 \Bigg(\frac{2}{q^2 (k+q)^2}
+ \frac{2}{(k+q)^2 (q+p)^2}\Bigg)
\Bigg]
-\nonumber\\
&& -\sum\limits_i c_i \int \frac{d^4k}{(2\pi)^4}\frac{d^4q}{(2\pi)^4}
\frac{e^2}{k^2 \Big(1+(-1)^n k^{2n}/\Lambda^{2n}\Big)}
\times\nonumber\\
&& \times
\Bigg[
\,V\partial^2 \Pi_{1/2} V \frac{2 \Big(q^2+M_i^2\Big)}{
\Big((k+q)^2-M_i^2\Big) \Big(q^2-M_i^2\Big)^2 \Big((q+p)^2-M_i^2\Big)}
-\nonumber\\
&& \qquad\qquad\qquad\qquad\quad\ \
- V^2 \frac{2 \Big(q^2+M_i^2\Big) \Big(q^2 + (q+p)^2\Big)
- 8 M_i^2 q^2}{\Big((k+q)^2-M_i^2\Big)
\Big(q^2-M_i^2\Big)^2 \Big((q+p)^2-M_i^2\Big) } \Bigg];\qquad
\\
\nonumber\\
\nonumber\\
\nonumber\\
\label{Two_Loop_Diagram4}
&&
\smash{\epsfxsize6.0truecm\epsfbox[190 390 490 890]{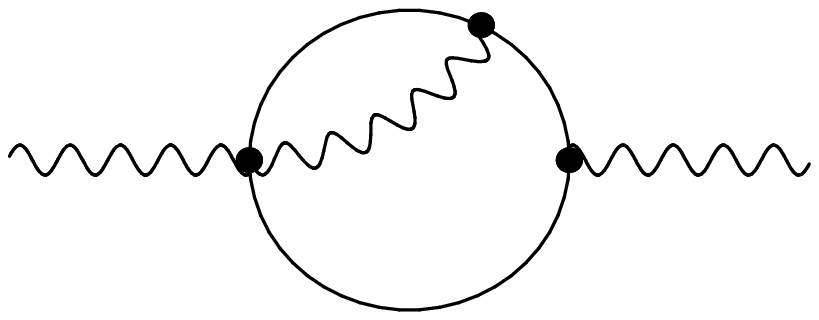}}
\hspace*{-1.2cm}
= \int \frac{d^4k}{(2\pi)^4}\frac{d^4q}{(2\pi)^4}
\frac{e^2}{k^2 \Big(1+(-1)^n k^{2n}/\Lambda^{2n}\Big)}
\times\vphantom{\int\limits_p}\nonumber\\
&& \times
\Bigg[
- V\partial^2\Pi_{1/2}V \frac{4}{(k+q)^2 q^2 (q+p)^2}
+ V^2 \Bigg(\frac{4}{q^2 (k+q)^2}
+ \frac{4}{(k+q)^2 (q+p)^2}\Bigg)
\Bigg]
-\vphantom{\int\limits^h}\nonumber\\
&& - \sum\limits_i c_i \int \frac{d^4k}{(2\pi)^4}\frac{d^4q}{(2\pi)^4}
\frac{e^2}{k^2 \Big(1+(-1)^n k^{2n}/\Lambda^{2n}\Big)}
\Bigg[
- V\partial^2\Pi_{1/2} V
\times\nonumber\\
&& \qquad\qquad\qquad\qquad\qquad\quad
\times
\frac{4}{\Big((k+q)^2-M_i^2\Big)
\Big(q^2-M_i^2\Big) \Big((q+p)^2-M_i^2\Big)}
+\nonumber\\
&& + V^2 \Bigg(\frac{4}{\Big(q^2-M_i^2\Big) \Big((k+q)^2-M_i^2\Big)}
+ \frac{4}{\Big((k+q)^2-M_i^2\Big) \Big((q+p)^2-M_i^2\Big)}\Bigg)
\Bigg];
\\
\nonumber\\
\nonumber\\
\nonumber\\
\label{Two_Loop_Diagram5}
&&
\smash{\epsfxsize6.0truecm\epsfbox[190 365 490 865]{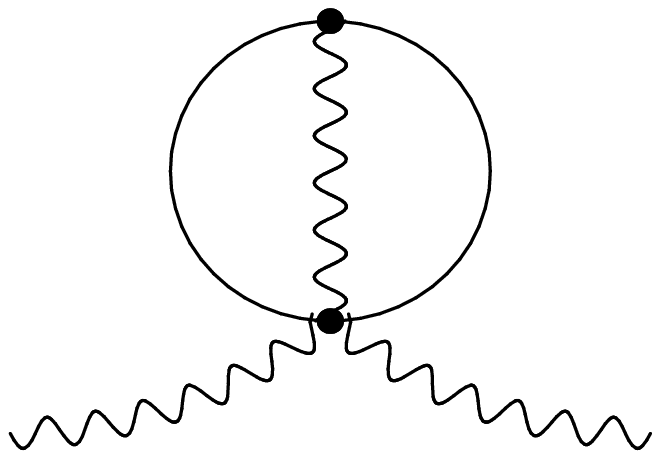}}
\hspace*{-2.0cm}
= \int \frac{d^4k}{(2\pi)^4}\frac{d^4q}{(2\pi)^4}\,
\frac{e^2}{k^2 \Big(1+(-1)^n k^{2n}/\Lambda^{2n}\Big)} V^2
\Bigg[- \frac{2}{q^2 (k+q)^2}
+\nonumber\\
&& \qquad\qquad\qquad\qquad\qquad\qquad\qquad\qquad\quad
+\sum\limits_i c_i\,
\frac{2}{\Big(q^2-M_i^2\Big) \Big((k+q)^2-M_i^2\Big)}
\Bigg];
\\
\nonumber\\
\nonumber\\
\label{Two_Loop_Diagram6}
&&
\smash{\epsfxsize6.0truecm\epsfbox[190 365 490 865]{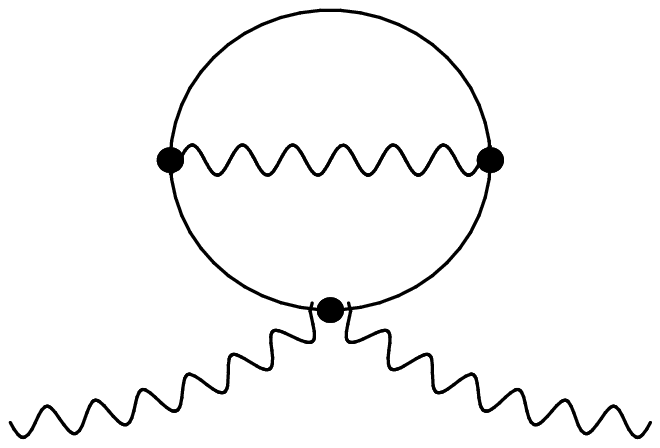}}
\hspace*{-2.0cm}
= \int \frac{d^4k}{(2\pi)^4}\frac{d^4q}{(2\pi)^4}\,
\frac{e^2}{k^2 \Big(1+(-1)^n k^{2n}/\Lambda^{2n}\Big)} V^2
\Bigg[\,\frac{2}{q^2 (k+q)^2}
-\nonumber\\
&& \qquad\qquad\qquad\qquad\qquad\qquad\qquad\qquad\ \
-\sum\limits_i c_i\,
\frac{2 \Big(q^2+M_i^2\Big)}{\Big(q^2-M_i^2\Big)^2 \Big((k+q)^2-M_i^2\Big)}
\Bigg];
\\
\nonumber\\
\nonumber\\
\label{Counterterms_Diagram1}
&&
\smash{\epsfxsize6.0truecm\epsfbox[190 390 490 890]{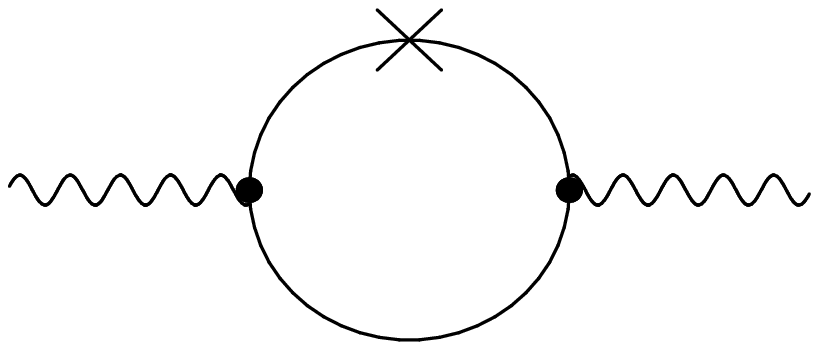}}
\hspace*{-1.2cm}
= -\frac{i e^2}{4\pi^2}\ln\frac{\Lambda}{\mu}
\int \frac{d^4k}{(2\pi)^4} \Bigg[
\, V\partial^2 \Pi_{1/2} V \frac{1}{k^2 (k+p)^2}
-\vphantom{\int\limits_p}\nonumber\\
&& - V^2 \Bigg(\frac{1}{k^2}+\frac{1}{(k+p)^2}\Bigg)
\Bigg]
+ \sum\limits_i c_i
\frac{i e^2}{4\pi^2}\ln\frac{\Lambda}{\mu}
\int \frac{d^4k}{(2\pi)^4} \Bigg[
\,V\partial^2\Pi_{1/2} V
\times\nonumber\\
&& \times
\frac{k^2+M_i^2}{(k^2-M_i^2)^2\Big((k+p)^2-M_i^2\Big)}
- V^2 \frac{(k^2+M_i^2)\Big(k^2 + (k+p)^2\Big)-4k^2 M_i^2}{
(k^2-M_i^2)^2\Big((k+p)^2-M_i^2\Big)}
\Bigg];
\\
\nonumber\\
\nonumber\\
\nonumber\\
\label{Counterterms_Diagram2}
&&
\smash{\epsfxsize6.0truecm\epsfbox[190 390 490 890]{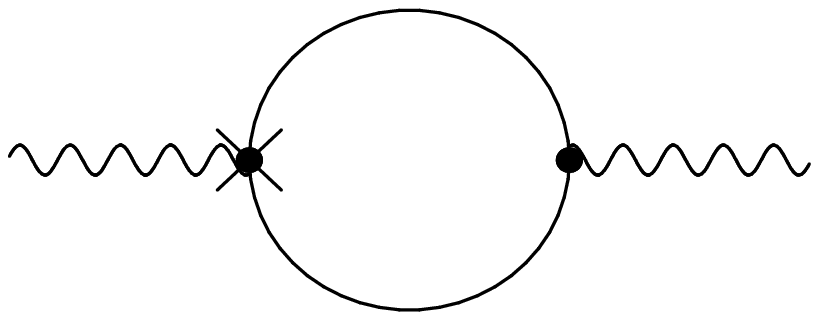}}
\hspace*{-1.2cm}
= \frac{i e^2}{4\pi^2}\ln\frac{\Lambda}{\mu}
\int \frac{d^4k}{(2\pi)^4} \Bigg[
\, V\partial^2\Pi_{1/2} V \frac{1}{k^2 (k+p)^2}
-\vphantom{\int\limits_p}\nonumber\\
&& - V^2 \Bigg(\frac{1}{k^2}+\frac{1}{(k+p)^2} \Bigg)
\Bigg]
- \sum\limits_i c_i
\frac{i e^2}{4\pi^2}\ln\frac{\Lambda}{\mu}
\int \frac{d^4k}{(2\pi)^4} \Bigg[
\,V\partial^2\Pi_{1/2}V
\times\nonumber\\
&& \times
\frac{1}{(k^2-M_i^2)\Big((k+p)^2-M_i^2\Big)}
- V^2 \Bigg(\frac{1}{k^2-M_i^2} + \frac{1}{(k+p)^2-M_i^2}\Bigg)
\Bigg];
\\
\nonumber\\
\nonumber\\
\nonumber\\
\label{Counterterms_Diagram3}
&& \smash{\epsfxsize6.0truecm\epsfbox[190 365 490 865]{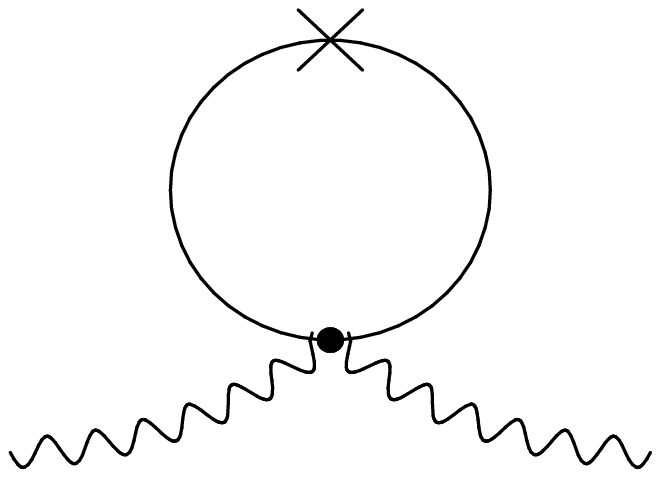}}
\hspace*{-2.0cm}
= \frac{i e^2}{4\pi^2}\ln\frac{\Lambda}{\mu}
\int \frac{d^4k}{(2\pi)^4} V^2 \Bigg(
-\frac{1}{k^2}+\sum\limits_i c_i\, \frac{k^2+M_i^2}{(k^2-M_i^2)^2}\Bigg);
\\
\nonumber\\
\nonumber\\
\nonumber\\
\label{Counterterms_Diagram4}
&&
\smash{\epsfxsize6.0truecm\epsfbox[190 365 490 865]{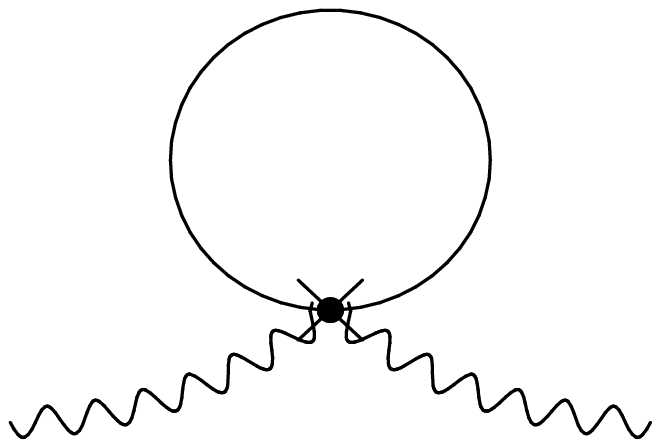}}
\hspace*{-2.0cm}
= \frac{i e^2}{4\pi^2}\ln\frac{\Lambda}{\mu}
\int \frac{d^4k}{(2\pi)^4} V^2 \Bigg(\frac{1}{k^2}
- \sum\limits_i c_i\,\frac{1}{k^2-M_i^2}\Bigg).
\\
\nonumber\\
\nonumber
\end{eqnarray}

Expressions for diagrams, giving nontrivial contributions to the
two-loop anomalous dimension are presented below. Again, in order
to obtain corresponding contributions to the effective action it
is necessary to add the factor

\begin{equation}
\int d^4\theta \frac{d^4p}{(2\pi)^4} \Big(\phi^*(p,\theta)\,\phi(-p,\theta)
+ \tilde\phi^*(p,\theta)\,\tilde\phi(-p,\theta)\Big).
\end{equation}

\begin{eqnarray}
&&
\smash{\epsfxsize6.0truecm\epsfbox[190 390 490 890]{dga1.eps}}
\hspace*{-1.2cm}
= \frac{i}{2}\int\frac{d^4k}{(2\pi)^4}
\frac{e^2}{k^2 \Big(1 + (-1)^n k^{2n}/\Lambda^{2n}\Big) (k+p)^2};
\\
\nonumber\\
\nonumber\\
\nonumber\\
&& \smash{\epsfxsize6.0truecm\epsfbox[190 390 490 890]{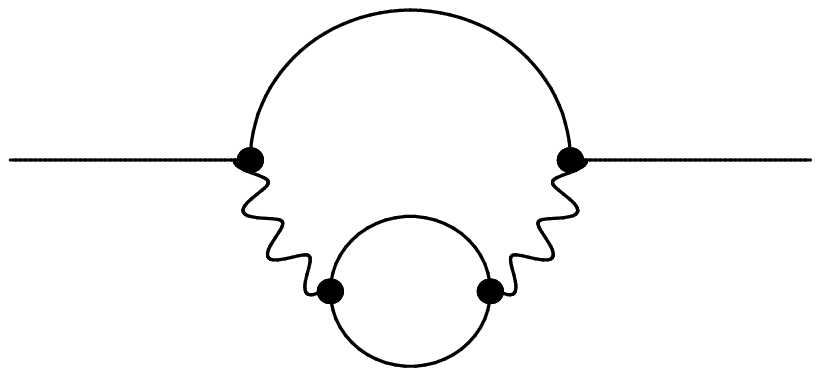}}
\hspace*{-1.2cm}
= - \int\frac{d^4k}{(2\pi)^4}\frac{d^4q}{(2\pi)^4}\frac{e^4}
{q^2 \Big(1+(-1)^n q^{2n}/\Lambda^{2n}\Big)^2 (q+p)^2}
\times\nonumber\\
&& \qquad\qquad\qquad\qquad\qquad\quad\
\times
\Bigg(\frac{1}{k^2 (k+q)^2} - \sum\limits_i c_i\,
\frac{1}{\Big(k^2-M_i^2\Big)\Big((k+q)^2-M_i^2\Big)}\Bigg)
+\nonumber\\
&&
+ 2\int\frac{d^4k}{(2\pi)^4}\frac{d^4q}{(2\pi)^4}\frac{e^4}
{q^4\Big(1+(-1)^n q^{2n}/\Lambda^{2n}\Big)^2 (q+p)^2}
\Bigg(\frac{1}{k^2} - \sum\limits_i c_i\, \frac{1}{k^2 - M_i^2}\Bigg)
+\nonumber\\
&&
+\int\frac{d^4k}{(2\pi)^4}\frac{d^4q}{(2\pi)^4}\frac{e^4}
{q^4 \Big(1+(-1)^n q^{2n}/\Lambda^{2n}\Big)^2}
\Bigg(\frac{1}{k^2 (k+q)^2}
-\\
&&\qquad\qquad\qquad\qquad\qquad\qquad\qquad\quad
- \sum\limits_i c_i\,
\frac{1}{\Big(k^2-M_i^2\Big)\Big((k+q)^2-M_i^2\Big)}\Bigg);
\nonumber\\
\nonumber\\
\nonumber\\
&& \smash{\epsfxsize6.0truecm\epsfbox[190 390 490 890]{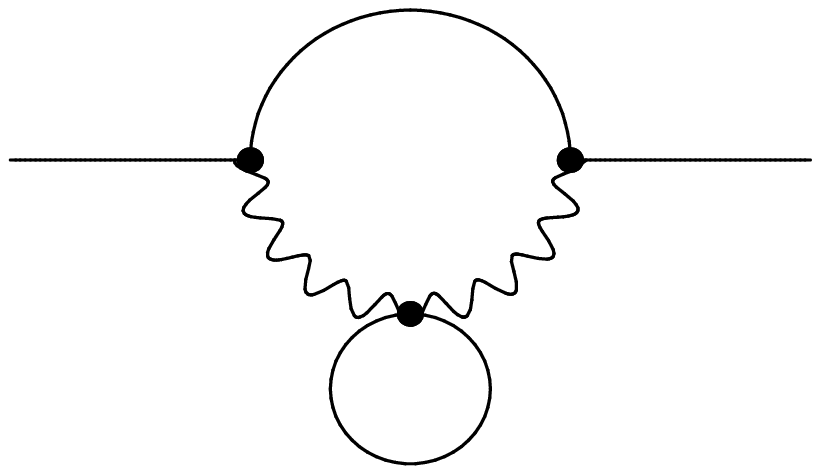}}
\hspace*{-1.2cm} =
- 2\int\frac{d^4k}{(2\pi)^4}\frac{d^4q}{(2\pi)^4}
\frac{e^4}
{q^4\Big(1+(-1)^n q^{2n}/\Lambda^{2n}\Big)^2 (q+p)^2}
\times\nonumber\\
&& \qquad\qquad\qquad\qquad\qquad\qquad
\times
\Bigg(\frac{1}{k^2} - \sum\limits_i c_i\, \frac{1}{k^2 - M_i^2}\Bigg);
\\
\nonumber\\
\nonumber\\
\nonumber\\
\nonumber\\
&&
\smash{\epsfxsize6.0truecm\epsfbox[190 365 490 865]{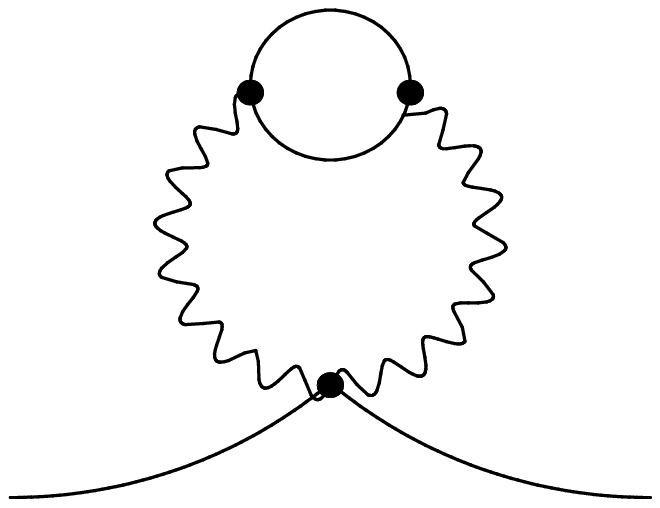}}
\hspace*{-2.0cm}
= - \int\frac{d^4k}{(2\pi)^4} \frac{d^4q}{(2\pi)^4}
\frac{e^4} {q^4 \Big(1+(-1)^n q^{2n}/\Lambda^{2n}\Big)^2}
\Bigg(\frac{1}{k^2 (k+q)^2}
-\nonumber\\
&& \qquad\qquad\qquad\qquad\qquad\qquad
- \sum\limits_i c_i\,
\frac{1}{\Big(k^2 - M_i^2\Big)\Big((k+q)^2 - M_i^2\Big)} \Bigg);
\\
\nonumber\\
\nonumber\\
&&
\smash{\epsfxsize6.0truecm\epsfbox[190 390 490 890]{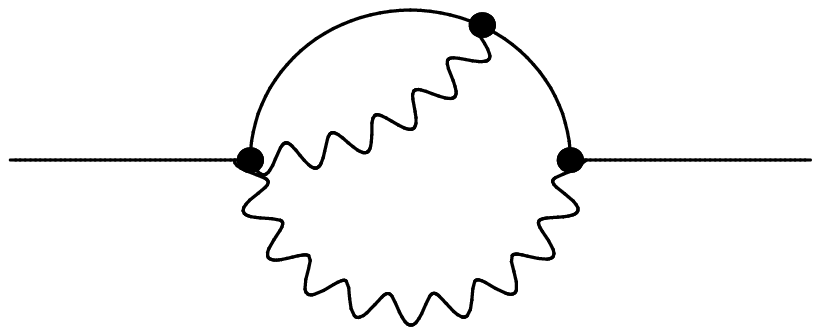}}
\hspace*{-1.2cm}
= - 2\int\frac{d^4k}{(2\pi)^4}\frac{d^4q}{(2\pi)^4}
\times \nonumber\\
\\
&& \times \frac{e^4}
{k^2\Big(1+(-1)^n k^{2n}/\Lambda^{2n}\Big)
q^2 \Big(1+(-1)^n q^{2n}/\Lambda^{2n}\Big)
(q+p)^2(k+q+p)^2 };
\nonumber\\
\nonumber\\
\nonumber\\
&& \smash{\epsfxsize6.0truecm\epsfbox[190 390 490 890]{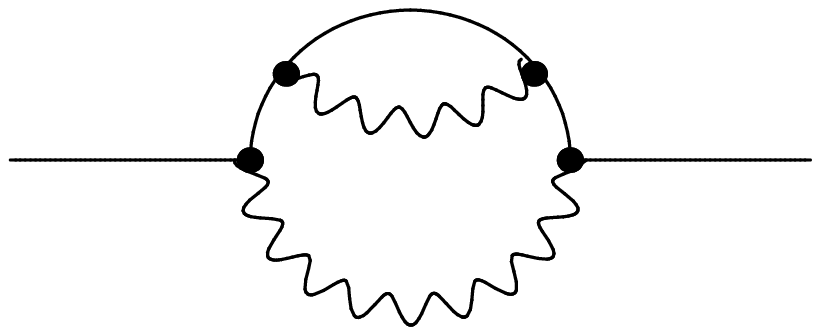}}
\hspace*{-1.2cm}
=\nonumber\\
\vphantom{\frac{1}{2}}\\
&&
= \int\frac{d^4k}{(2\pi)^4}\frac{d^4q}{(2\pi)^4}\frac{e^4}
{k^2 \Big(1+(-1)^n k^{2n}/\Lambda^{2n}\Big)
q^2 \Big(1+(-1)^n q^{2n}/\Lambda^{2n}\Big) (q+p)^2(k+q+p)^2 };
\nonumber\\
\nonumber\\
\nonumber\\
&&
\nonumber\\
&& \smash{\epsfxsize5.8truecm\epsfbox[190 390 490 890]{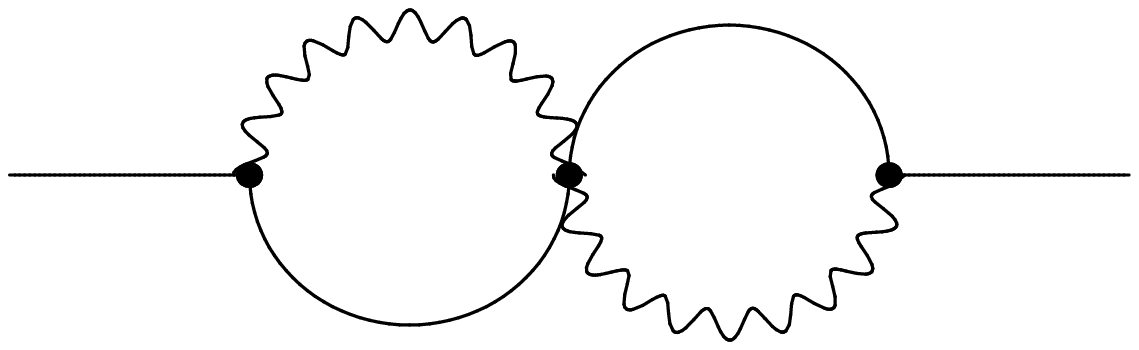}}
\hspace*{-0.2cm}
=\nonumber\\
\vphantom{\frac{1}{2}}\\
&&
= - \int\frac{d^4k}{(2\pi)^4}\frac{d^4q}{(2\pi)^4}\frac{e^4}
{k^2\Big(1+(-1)^n k^{2n}/\Lambda^{2n}\Big)
q^2 \Big(1+(-1)^n q^{2n}/\Lambda^{2n}\Big) (q+p)^2(k+p)^2};
\nonumber\\
\nonumber\\
\nonumber\\
\nonumber\\
&& \smash{\epsfxsize6.0truecm\epsfbox[190 390 490 890]{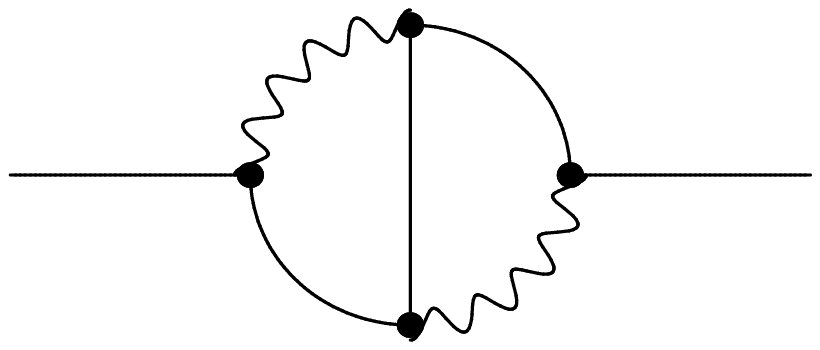}}
\hspace*{-1.2cm}
= \int\frac{d^4k}{(2\pi)^4}\frac{d^4q}{(2\pi)^4}\times\nonumber\\
\\
&&
\times
\frac{e^4 (q+k+2p)^2}
{k^2 \Big(1+(-1)^n k^{2n}/\Lambda^{2n}\Big) q^2
\Big(1+(-1)^n q^{2n}/\Lambda^{2n}\Big) (k+q+p)^2 (q+p)^2 (k+p)^2};
\nonumber\\
\nonumber\\
\nonumber\\
\nonumber\\
&& \smash{\epsfxsize6.0truecm\epsfbox[190 390 490 890]{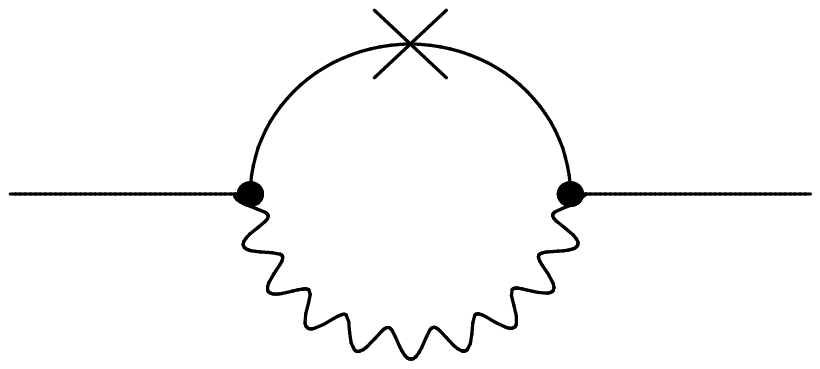}}
\hspace*{-1.2cm}
= -\frac{i}{8\pi^2}\ln\frac{\Lambda}{\mu}\int\frac{d^4k}{(2\pi)^4}
\frac{e^4}{k^2\Big(1+(-1)^n k^{2n}/\Lambda^{2n}\Big)(k+p)^2};
\nonumber\\
\\
\nonumber\\
\nonumber\\
&& \smash{\epsfxsize6.0truecm\epsfbox[190 390 490 890]{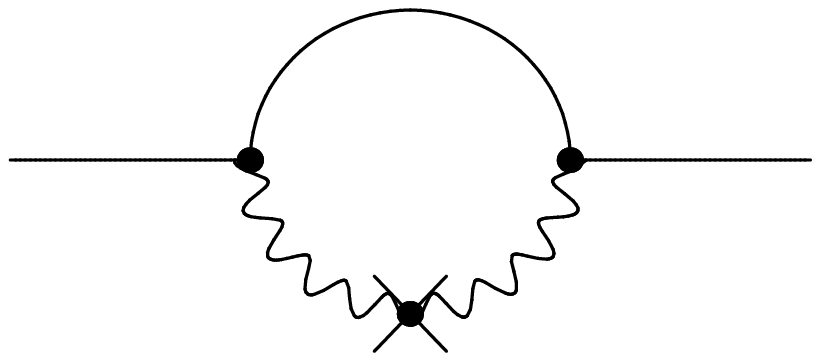}}
\hspace*{-1.2cm}
= - \frac{i}{8\pi^2}\ln\frac{\Lambda}{\mu}
\int\frac{d^4k}{(2\pi)^4}
\frac{e^4}{k^4\Big(1+(-1)^n k^{2n}/\Lambda^{2n}\Big)}
+\\
&& \qquad\qquad\qquad\qquad\qquad\qquad
+ \frac{i}{8\pi^2}\ln\frac{\Lambda}{\mu}\int\frac{d^4k}{(2\pi)^4}
\frac{e^4}{k^2\Big(1+(-1)^n k^{2n}/\Lambda^{2n}\Big)(k+p)^2};
\nonumber\\
\nonumber\\
\nonumber\\
&&
\smash{\epsfxsize6.0truecm\epsfbox[190 390 490 890]{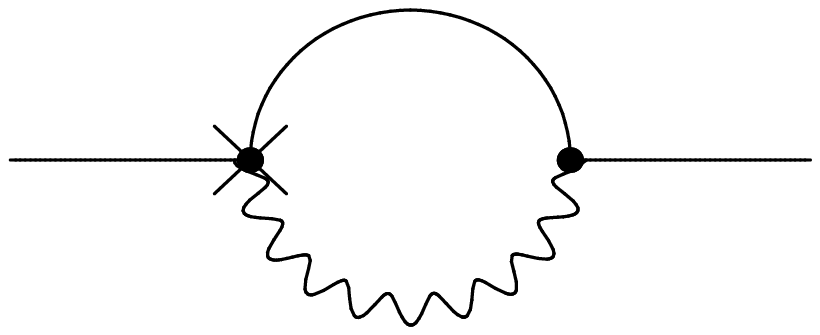}}
\hspace*{-1.2cm}
= \frac{i}{4\pi^2}\ln\frac{\Lambda}{\mu}
\int\frac{d^4k}{(2\pi)^4}\frac{e^4}{k^2
\Big(1+(-1)^n k^{2n}/\Lambda^{2n}\Big) (k+p)^2};
\nonumber\\
\\
\nonumber\\
\nonumber\\
\nonumber\\
&&
\smash{\epsfxsize6.0truecm\epsfbox[190 365 490 865]{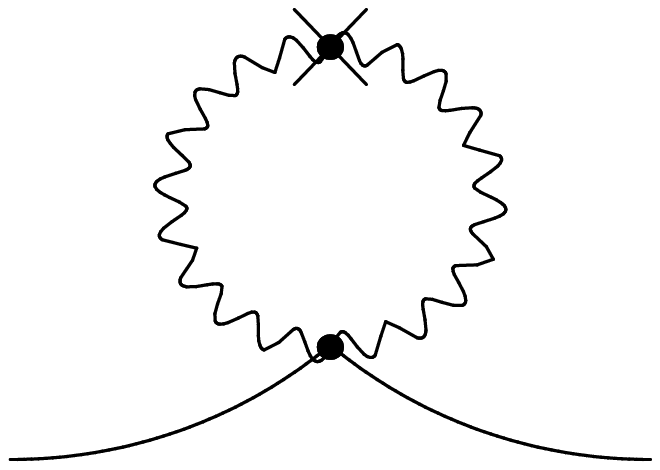}}
\hspace*{-2.0cm}
= \frac{i}{8\pi^2} \ln\frac{\Lambda}{\mu}
\int\frac{d^4k}{(2\pi)^4}
\frac{e^4}{k^4\Big(1+ (-1)^n k^{2n}/\Lambda^{2n}\Big)}.
\\
\nonumber
\end{eqnarray}

%%%%%%%%%%%%%%%%%%%%%%%%%%%%%%%%%%%%%%%%%%%%%%%%%%%%%%%%%%%%%%%%%%%%%%%%%%%

\section{Calculation of integrals, regularized by higher derivatives.}
\hspace{\parindent}
\label{Appendix_Integrals}

Let us calculate divergent parts of integrals, which were encountered
in equations (\ref{Two_Loop_Effective_Action_V}) and
(\ref{Two_Loop_Effective_Action_Phi}). First it is necessary to perform
the Wick rotation. After it the integrals, which are present in
equations (\ref{Two_Loop_Effective_Action_V}) and
(\ref{Two_Loop_Effective_Action_Phi}) will be proportional to

\begin{eqnarray}\label{Integrals_Definition}
&& I_1 = \int d^4 k \frac{1}{k^2 (k+p)^2
\Big(1+ k^{2n}/\Lambda^{2n}\Big)};\nonumber\\
&& I_2 = \int d^4k\, \frac{1}{k^2 (k+p)^2}
- \sum\limits_i c_i
\int d^4k\,\frac{1}{(k^2+M_i^2)\Big((k+p)^2+M_i^2\Big)};
\nonumber\\
&& I_3 = \int d^4k\,d^4q
\frac{(k+p+q)^2+q^2-k^2-p^2}{k^2 \Big(1 + k^{2n}/\Lambda^{2n}\Big)
(k+q)^2 (k+p+q)^2 q^2 (q+p)^2};\nonumber\\
&& I_4 = \int d^4k\,\frac{M^2}{(k^2+M^2)^2 \Big((k+p)^2+M^2\Big)};
\nonumber\\
&& I_5 = \int d^4k\,d^4q\,
\frac{1}{k^2 (k+p)^2 q^2 (q+p)^2 \Big(1+k^{2n}/\Lambda^{2n}\Big)
\Big(1+ q^{2n}/\Lambda^{2n}\Big)}=I_1^2;\nonumber\\
&& I_6 = \int d^4k\,d^4q\,
\frac{(k+q+2p)^2}{k^2 (k+p)^2 q^2 (q+p)^2 (k+q+p)^2
\Big(1+k^{2n}/\Lambda^{2n}\Big)
\Big(1+ q^{2n}/\Lambda^{2n}\Big)};\qquad\nonumber\\
&& I_7 = \int d^4k\,d^4q\,
\frac{1}{\displaystyle k^2 q^2 (q+p)^2 (k+q+p)^2
\Big(1+k^{2n}/\Lambda^{2n}\Big)
\Big(1+ q^{2n}/\Lambda^{2n}\Big)};\nonumber\\
&& I_8 = \int d^4q
\,\frac{1}{q^2 (q+p)^2 \Big(1+ q^{2n}/\Lambda^{2n}\Big)^2}\, I_2(q/M);
\nonumber\\
&& I_9 = \int \frac{d^4k}{(2\pi)^4}\frac{d^4q}{(2\pi)^4}\,
\frac{1}{k^2 \Big(1 + k^{2n}/\Lambda^{2n}\Big)}
\times\nonumber\\
&& \qquad\qquad
\times
\frac{(k+p+q)^2+q^2-k^2-p^2}{
\Big((k+q)^2+M^2\Big) \Big((k+p+q)^2+M^2\Big) \Big(q^2+M^2\Big)
\Big((q+p)^2+M^2\Big)};\nonumber\\
&& I_{10} = \int \frac{d^4k}{(2\pi)^4}\frac{d^4q}{(2\pi)^4}\,
\frac{1}{k^2 \Big(1 + k^{2n}/\Lambda^{2n}\Big)}
\times\nonumber\\
&& \qquad\qquad\qquad\qquad\qquad
\times
\frac{M^2}{\Big((k+q)^2+M^2\Big) \Big(q^2+M^2\Big)^2
\Big((q+p)^2+M^2\Big)}.
\end{eqnarray}

In order to calculate integral $I_1$ it is possible to use
four-dimensional spherical coordinates

\begin{eqnarray}
&& k_1 = k\sin\theta_3\sin\theta_2\sin\theta_1;\nonumber\\
&& k_2 = k\sin\theta_3\sin\theta_2\cos\theta_1;\nonumber\\
&& k_3 = k\sin\theta_3\cos\theta_2;\nonumber\\
&& k_4 = k\cos\theta_3.\nonumber\\
\end{eqnarray}

\noindent
and direct fourth axis along $p^\mu$, so that the integrand will depend
only on $\theta_3$ and

\begin{eqnarray}
&& \int d^4k = \int\limits_{0}^\infty k^3 dk\,
\int\limits_0^\pi d\theta_3\, \sin^2\theta_3
\int\limits_0^\pi d\theta_2\, \sin\theta_2
\int\limits_0^{2\pi} d\theta_1 =
4\pi \int\limits_{0}^\infty k^3 dk\,
\int\limits_0^\pi d\theta_3\, \sin^2\theta_3
=\qquad\nonumber\\
&&= \Big[x=\cos\theta_3\Big] =
4\pi \int\limits_{0}^\infty k^3 dk\,
\int\limits_{-1}^1 dx\, \sqrt{1-x^2}.
\end{eqnarray}

\begin{figure}[h]

\hspace*{3cm}
\begin{picture}(0,0)(0,0)
\put(7.5,-3.9){$\mbox{Re}(x)$}
\put(4.8,-0.9){$\mbox{Im}(x)$}
\put(3.2,-4.1){$-1$}
\put(5.6,-4.1){$1$}
\put(2,-3.9){$x_0$}
\end{picture}

\hspace*{3cm}
\epsfxsize9.0truecm\epsfbox{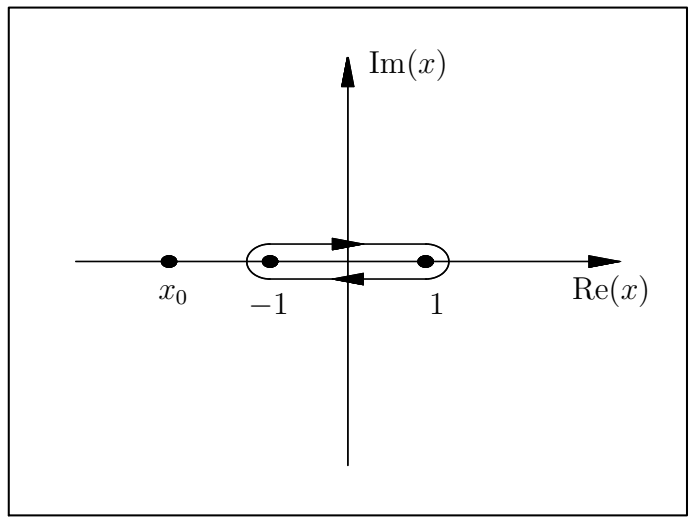}

\caption{Contour $C$ for calculation of integral over $x$.}
\label{Figure_Contour}
\end{figure}

\noindent
Taking into account, that $k^\mu p_\mu = kp\cos\theta_3 = kpx$, the
integral can be written as

\begin{eqnarray}
&& \int d^4k \frac{1}{k^2 (k+p)^2 \Big(1 + k^{2n}/\Lambda^{2n}\Big)}
=\nonumber\\
&&\qquad\qquad
= 4\pi \int\limits_0^\infty k\, dk\,
\int\limits_{-1}^{1} dx\,\frac{\sqrt{1-x^2}}{(k^2+2kpx+p^2)
\Big(1+k^{2n}/\Lambda^{2n}\Big)}
=\nonumber\\
&& \qquad\qquad\qquad\qquad
= 2\pi  \int\limits_0^\infty k\, dk\,
\oint\limits_C dx\,\frac{\sqrt{1-x^2}}{(k^2+2kpx+p^2)
\Big(1+k^{2n}/\Lambda^{2n}\Big)},\qquad
\end{eqnarray}

\noindent
where the contour $C$ is presented at Figure \ref{Figure_Contour}. The
integrand here has singularities at branch points $x = \pm 1$, a pole
$x=\infty$ and a pole

\begin{equation}
x_0 = -\frac{k^2+p^2}{2kp}.
\end{equation}

\noindent
Then it is easy to see, that

\begin{eqnarray}
&& \oint\limits_C dx\,\frac{\sqrt{1-x^2}}{k^2+2kpx+p^2} =
2\pi i\, \mbox{Res}\Bigg(\frac{\sqrt{1-x^2}}{k^2+2kpx+p^2},
\,x=\infty\Bigg)
-\nonumber\\
&& - 2\pi i\,\mbox{Res}\Bigg(\frac{\sqrt{1-x^2}}{k^2+2kpx+p^2},
\,x=x_0\Bigg)
= 2\pi i\Bigg(-i\frac{k^2+p^2}{4 k^2p^2}
+ i \frac{|k^2-p^2|}{4k^2p^2}\Bigg).\qquad
\end{eqnarray}

\noindent
Therefore, the integral over angles is reduced to

\begin{eqnarray}\label{Angle_Integral}
\oint dx\,\frac{\sqrt{1-x^2}}{k^2+2kpx+p^2} =
\left\{
\begin{array}{l}
{\displaystyle \frac{\pi}{k^2},\quad k\ge p;}\\
\\
{\displaystyle \frac{\pi}{p^2},\quad p\ge k}
\end{array}
\right.
\end{eqnarray}

\noindent
and finally

\begin{eqnarray}
&& I_1 = 2\pi^2 \int\limits_0^p dk\,\frac{k}{p^2}\,
\frac{1}{\Big(1+k^{2n}/\Lambda^{2n}\Big)}
+ 2\pi^2 \int\limits_p^\infty dk\,\frac{1}{k}\,
\frac{1}{\Big(1+k^{2n}/\Lambda^{2n}\Big)}
=\qquad\nonumber\\
&& = \pi^2 + o(1)
+ \frac{\pi^2}{n} \ln\frac{\Lambda^{2n}+p^{2n}}{p^{2n}}
= 2 \pi^2 \Big(\ln\frac{\Lambda}{p} + \frac{1}{2}\Big) + o(1).
\end{eqnarray}

Integral $I_2$ can be calculated using standard methods \cite{Ramond}.
First, using an identity

\begin{equation}
\frac{1}{ab} = \int\limits_0^1 dy\,\frac{1}{\Big(ay + b(1-y)\Big)^2},
\end{equation}

\noindent
it can be written as

\begin{equation}\label{Rewritten_I2}
I_2 = \int\limits_0^1 dy \int d^4k\,\Bigg(
\frac{1}{\Big(k^2 + 2kp y + y p^2\Big)^2}
-\sum\limits_i c_i \frac{1}{\Big(k^2+2kp y + y p^2 + M_i^2\Big)^2}\Bigg).
\end{equation}

\noindent
Each of these integrals diverges, but their difference is finite.
Therefore, to simplify calculations it is convenient to use an
auxiliary regularization, for example, the dimensional regularization.
Then the integrals in equation (\ref{Rewritten_I2}) can be easily
taken:

\begin{eqnarray}\label{I2_Result}
&& I_2 = \lim\limits_{D\to 4} \pi^2 \int\limits_0^1 dy\,
\frac{\Gamma(2-D/2)}{\Gamma(2)} \Bigg(\Big(y (1-y)\,p^2\Big)^{D/2-2}
-\nonumber\\
&& \qquad\qquad\qquad\qquad\qquad\qquad\qquad\quad
- \sum\limits_i c_i \Big(y (1-y)\,p^2 + M_i^2\Big)^{D/2-2}\Bigg)
=\nonumber\\
&&
= \pi^2 \sum\limits_i c_i
\int\limits_0^1 dy\,\ln\Bigg(1+\frac{M_i^2}{y(1-y)\,p^2}\Bigg)
=\\
&& \qquad\qquad\qquad\qquad
= 2\pi^2 \sum\limits_i c_i
\Bigg(\ln\frac{M_i}{p} + \sqrt{1+\frac{4M_i^2}{p^2}}
\mbox{arctanh}\sqrt{\frac{p^2}{4M_i^2 + p^2}}\Bigg).\qquad\nonumber
\end{eqnarray}

\noindent
where we take into account, that $\sum\limits_i c_i = 1$.

To calculate divergent part of the integral $I_3$ note, that
$I_3 = I_3(p/\Lambda)$ and due to the logarithmical divergence

\begin{equation}\label{I2_Expansion}
I_3 = a_1 \ln^2\frac{\Lambda}{p} + a_2 \ln\frac{\Lambda}{p}
+ \sum\limits_{i=0}^\infty b_i \Bigg(\frac{p^2}{\Lambda^2}\Bigg)^i.
\end{equation}

\noindent
If $a_1 = 0$, then it is possible to find

\begin{equation}\label{A2_Expression}
a_2 = \lim\limits_{p\to 0}\,\frac{d I_3}{d\ln\Lambda}
= \int d^4k\,d^4q\,
\frac{4n\,k^{2n-2}}{\Lambda^{2n} \Big(1+k^{2n}/\Lambda^{2n}\Big)^2}
\,\frac{q^2 + k_\mu q_\mu}{(k+q)^4 q^4}.
\end{equation}

\noindent
If this limit does not exist, then $a_1\ne 0$. The integral in the
right hand side of equation (\ref{A2_Expression}) can be taken, using
four-dimensional spherical coordinates:

\begin{eqnarray}\label{For_I3}
&& \int d^4q\,\frac{q^2 + k_\mu q_\mu}{(k+q)^4 q^4}
= 4\pi \int\limits_0^\infty dq \int\limits_{-1}^1 dx\,
\frac{(q + k x)\sqrt{1-x^2}}{(k^2 + 2kqx + q^2)^2}
=\nonumber\\
&& \qquad
=  - 2\pi \int\limits_{-1}^1 dx \int\limits_0^\infty dq\,
\frac{d}{d q} \frac{\sqrt{1-x^2}}{(k^2 + 2kqx + q^2)}
= \frac{2\pi}{k^2} \int\limits_{-1}^1 dx\,\sqrt{1-x^2}
= \frac{\pi^2}{k^2},\qquad
\end{eqnarray}

\noindent
so that

\begin{equation}
a_2 = 4n\pi^2 \int d^4k\,
\frac{k^{2n-4}}{\Lambda^{2n} \Big(1+k^{2n}/\Lambda^{2n}\Big)^2} = 4\pi^4,
\end{equation}

\noindent
Therefore, from equation (\ref{I2_Expansion}) we conclude, that

\begin{equation}
I_3 = 4\pi^4 \ln\frac{\Lambda}{p} + O(1).
\end{equation}

In order to calculate integral $I_4$ let us note, that

\begin{equation}
\int \frac{d^4k}{(2\pi)^4}\,\frac{M^2}{(k^2+M^2)^2 \Big((k+p)^2+M^2\Big)}
= f(p/M).
\end{equation}

\noindent
Therefore, instead of taking the limit $M\to \infty$ it is possible to
take the limit $p\to 0$, so that

\begin{eqnarray}
&& I_4 = \int d^4k\,\frac{M^2}{(k^2+M^2)^3} + o(1) = \frac{\pi^2}{2}
+ o(1).
\end{eqnarray}

Divergent part of integral $I_5$ can be also easily calculated, because

\begin{equation}
I_5 = I_1^2
= 4\pi^4\Bigg(\ln^2 \frac{\Lambda}{p} + \ln\frac{\Lambda}{p}\Bigg) + O(1).
\end{equation}

To find a divergent part of $I_6$ let us consider

\begin{eqnarray}
&& \lim\limits_{p\to 0}\,\Lambda \frac{d}{d\Lambda} \Big(I_5 - I_6\Big)
=\nonumber\\
&& = \lim\limits_{p\to 0}\,\Lambda \frac{d}{d\Lambda}\int d^4k\,d^4q\,
\Bigg(1 - \frac{(k+q+2p)^2}{(k+q+p)^2} \Bigg)
\times\nonumber\\
&& \qquad\qquad\qquad\qquad
\times
\frac{1}{k^2 (k+p)^2 q^2 (q+p)^2 \Big(1+k^{2n}/\Lambda^{2n}\Big)
\Big(1+ q^{2n}/\Lambda^{2n}\Big)}
=\qquad\nonumber\\
&& = \lim\limits_{p\to 0} \int d^4k\,d^4q\,
\frac{- 2(k+q)p - 3p^2}{(k+q+p)^2}
\times\nonumber\\
&& \qquad\qquad\qquad\qquad
\times
\frac{4n q^{2n}/\Lambda^{2n}}{k^2 (k+p)^2 q^2 (q+p)^2
\Big(1+k^{2n}/\Lambda^{2n}\Big)
\Big(1+ q^{2n}/\Lambda^{2n}\Big)^2} = 0.\qquad
\end{eqnarray}

\noindent
(It is important to note, that all integrals here are convergent.)
Therefore,

\begin{equation}
I_6 = I_1^2 + O(1)
= 4\pi^4\Bigg(\ln^2 \frac{\Lambda}{p} + \ln\frac{\Lambda}{p}\Bigg) + O(1).
\end{equation}

A divergent part of $I_7$ can be calculated similarly:

\begin{eqnarray}
&& \lim\limits_{p\to 0}\,\Lambda \frac{d}{d\Lambda} \Big(I_5 - 2 I_7\Big)
=\nonumber\\
&& = \lim\limits_{p\to 0}\,\Lambda \frac{d}{d\Lambda}\int d^4k\,d^4q\,
\Bigg(\frac{1}{(k+p)^2} - \frac{2}{(k+q+p)^2} \Bigg)
\times\nonumber\\
&& \qquad\qquad\qquad\qquad\qquad\qquad
\times
\frac{1}{k^2 q^2 (q+p)^2 \Big(1+k^{2n}/\Lambda^{2n}\Big)
\Big(1+ q^{2n}/\Lambda^{2n}\Big)}
=\qquad\nonumber\\
&& = \lim\limits_{p\to 0}\,\Lambda \frac{d}{d\Lambda}\int d^4k\,d^4q\,
\frac{q^2 + 2 (k+p)q - (k+p)^2}{(k+p)^2 (k+q+p)^2}
\times\nonumber\\
&& \qquad\qquad\qquad\qquad\qquad\qquad
\times
\frac{1}{k^2 q^2 (q+p)^2 \Big(1+k^{2n}/\Lambda^{2n}\Big)
\Big(1+ q^{2n}/\Lambda^{2n}\Big)}
=\nonumber\\
&& = \lim\limits_{p\to 0}\,\Lambda \frac{d}{d\Lambda}\int d^4k\,d^4q\,
\frac{q^2 + 2 (k+p)q - (q+p)^2}{(k+p)^2 (k+q+p)^2}
\times\nonumber\\
&& \qquad\qquad\qquad\qquad\qquad\qquad
\times
\frac{1}{k^2 q^2 (q+p)^2 \Big(1+k^{2n}/\Lambda^{2n}\Big)
\Big(1+ q^{2n}/\Lambda^{2n}\Big)}
=\nonumber\\
&& = \lim\limits_{p\to 0}\,\Lambda \frac{d}{d\Lambda}\int d^4k\,d^4q\,
\frac{2 kq - p^2}{(k+p)^2 (k+q+p)^2}
\times\nonumber\\
&& \qquad\qquad\qquad\qquad\qquad\qquad
\times
\frac{1}{k^2 q^2 (q+p)^2 \Big(1+k^{2n}/\Lambda^{2n}\Big)
\Big(1+ q^{2n}/\Lambda^{2n}\Big)}
=\nonumber\\
&& = \int d^4k\,d^4q\,
\frac{2kq}{k^4 q^4 (k+q)^2}
\,\Lambda \frac{d}{d\Lambda}\Bigg[
\frac{1}{\Big(1+k^{2n}/\Lambda^{2n}\Big)
\Big(1+ q^{2n}/\Lambda^{2n}\Big)}\Bigg]
=\nonumber\\
&& \qquad\qquad\qquad\quad
= 8n\int d^4q\,d^4k\,
\frac{kq}{k^4 q^4 (k+q)^2}
\frac{q^{2n}/\Lambda^{2n}}{\Big(1+k^{2n}/\Lambda^{2n}\Big)
\Big(1+ q^{2n}/\Lambda^{2n}\Big)^2}.\qquad
\end{eqnarray}

\noindent
To calculate this integral we again use four-dimensional spherical
coordinates and direct fourth axis along $q^\mu$. Then similar to the
case, considered above, the integral over angles is reduced to

\begin{eqnarray}
&& 4\pi\int\limits_{-1}^{1} dx\,\frac{x \sqrt{1-x^2}}{k^2+2kqx+q^2}
= 2\pi \oint\limits_C dx\,\frac{x\sqrt{1-x^2}}{k^2+2kqx+q^2}
=\nonumber\\
&& = 4\pi^2 i\, \mbox{Res}\Bigg(\frac{x\sqrt{1-x^2}}{k^2+2kqx+q^2},
\,x=\infty\Bigg)
- 4\pi^2 i\,\mbox{Res}\Bigg(\frac{x\sqrt{1-x^2}}{k^2+2kqx+q^2},
\,x=x_0\Bigg)
=\qquad\nonumber\\
&& = 4\pi^2 i\Bigg(- \frac{i}{4 kq} + \frac{i (k^2+q^2)^2}{8 k^3 q^3}
- \frac{i|k^2-q^2| (k^2 + q^2)}{8 k^3 q^3}\Bigg),
\end{eqnarray}

\noindent
so that

\begin{eqnarray}
2\pi \oint dx\,\frac{x\sqrt{1-x^2}}{k^2+2kqx+q^2} =
\left\{
\begin{array}{l}
{\displaystyle - \frac{\pi^2 q}{k^3},\quad k\ge q;}\\
\\
{\displaystyle - \frac{\pi^2 k}{q^3},\quad q\ge k.}
\end{array}
\right.
\end{eqnarray}

\noindent
Therefore,

\begin{eqnarray}
&& \lim\limits_{p\to 0}\,\Lambda \frac{d}{d\Lambda} \Big(I_5 - 2 I_7\Big)
= - 16n\,\pi^4 \int\limits_0^\infty dq\,
\frac{q^{2n}/\Lambda^{2n}}{\Big(1+q^{2n}/\Lambda^{2n}\Big)^2}
\times\nonumber\\
&&\qquad\qquad\qquad\quad\ \ \times
\Bigg(\int\limits_q^\infty dk\,
\frac{q}{k^3 \Big(1+ k^{2n}/\Lambda^{2n}\Big)}
+ \int\limits_0^q dk\,
\frac{k}{q^3 \Big(1+ k^{2n}/\Lambda^{2n}\Big)}
\Bigg)
=\nonumber\\
&&
= - 4n\,\pi^4 \int\limits_0^\infty dx\,
\frac{x^{n}}{(1+x^{n})^2}
\int\limits_0^{1/x}
\frac{dy}{1+ y^{-n}}
- 4n\,\pi^4 \int\limits_0^\infty dx\,
\frac{x^{n}}{(1+x^{n})^2}
\int\limits_0^{1/x}
\frac{dy}{1+ y^{n}}
=\nonumber\\
&&
= - 4n\,\pi^4 \int\limits_0^\infty dx\,
\frac{x^{n-1}}{(1+x^{n})^2}
= - 4\pi^4
\end{eqnarray}

\noindent
and finally

\begin{equation}
I_7 = \frac{1}{2} I_1^2 + 2\pi^4 \ln\frac{\Lambda}{p} + O(1)
= 2\pi^4\Bigg(\ln^2 \frac{\Lambda}{p} + 2\ln\frac{\Lambda}{p}\Bigg) + O(1).
\end{equation}

Using equation (\ref{I2_Result}) integral $I_8$ can be written as

\begin{eqnarray}\label{Integrals_DK_Integral}
&& I_8 = 2 \pi^2 \sum\limits_i c_i \int d^4q\, \frac{1}{q^2 (q+p)^2
\Big(1+q^{2n}/\Lambda^{2n}\Big)^2}
\times\nonumber\\
&& \qquad\qquad\qquad\qquad\qquad\quad
\times
\Bigg(\ln \frac{M_i}{q}
+ \sqrt{1+\frac{4M_i^2}{q^2}}
\mbox{arctanh}\sqrt{\frac{q^2}{4M_i^2 + q^2}}\Bigg),\qquad
\end{eqnarray}

\noindent
where $M_i = a_i\Lambda$, $a_i$ being constants. To calculate the
divergent part of this integral let us consider first an integral

\begin{equation}
I_f \equiv \int d^4q \frac{1}{q^2 (q+p)^2} f\Big(\Lambda/q\Big)
= I_f\Big(\Lambda/p\Big),
\end{equation}

\noindent
where $f$ is a function. Differentiating $I_f$ over $\ln \Lambda$
and setting then $p=0$, we obtain, that

\begin{eqnarray}\label{Integrals_DIF}
&& \Lambda\frac{d I_f}{d\Lambda} \Bigg|_{p=0}
= \int d^4q \frac{1}{q^4}\,\Lambda\frac{d}{d\Lambda} f\Big(\Lambda/q\Big)
= - \int d^4q \frac{1}{q^3}\,\frac{d}{dq} f\Big(\Lambda/q\Big)
=\nonumber\\
&& \qquad\qquad\qquad\qquad\qquad\quad
= - 2\pi^2 \int\limits_0^\infty dq\, \frac{d}{dq} f\Big(\Lambda/q\Big)
= 2\pi^2 \Big(f(\infty) - f(0)\Big).\qquad\quad
\end{eqnarray}

\noindent
So, if the values $f(\infty)$ and $f(0)$ are finite, then

\begin{equation}\label{Integrals_IF}
I_f = 2\pi^2 \Big(f(\infty) - f(0)\Big)\ln\frac{\Lambda}{p} + O(1).
\end{equation}

\noindent
If the function $f$ is taken to be

\begin{equation}
f\Big(\Lambda/q\Big) = \sum\limits_i c_i
\frac{2\pi^2}{\Big(1+q^{2n}/\Lambda^{2n}\Big)^2}
\sqrt{1+\frac{4M_i^2}{q^2}}\,\mbox{arctanh}\sqrt{\frac{q^2}{4M_i^2 + q^2}},
\end{equation}

\noindent
then from equation (\ref{Integrals_IF}) we conclude, that

\begin{eqnarray}\label{Integrals_Atan_Integral}
&& 2\pi^2 \sum\limits_i c_i \int d^4q \frac{1}{q^2 (q+p)^2
\Big(1+q^{2n}/\Lambda^{2n}\Big)^2}
\sqrt{1+\frac{4M_i^2}{q^2}} \mbox{arctanh}\sqrt{\frac{q^2}{4M_i^2 + q^2}}
=\qquad\nonumber\\
&&
= 4\pi^4 \sum\limits_i c_i \ln\frac{\Lambda}{p} + O(1)
= 4\pi^4 \ln\frac{\Lambda}{p} + O(1).\qquad
\end{eqnarray}

\noindent
However, it is impossible to substitute in equation (\ref{Integrals_IF})

\begin{equation}
f\Big(\Lambda/q\Big)
= \frac{2\pi^2}{\Big(1+q^{2n}/\Lambda^{2n}\Big)^2}
\sum\limits_i c_i \ln \frac{M_i}{q}
\end{equation}

\noindent
because $f(\infty)$ does not exist. Nevertheless, the function $f$
can chosen in following form:

\begin{eqnarray}
&& f\Big(\Lambda/q\Big)
= \Lambda\frac{d}{d\Lambda}
\Bigg(\frac{2\pi^2}{\Big(1+q^{2n}/\Lambda^{2n}\Big)^2}
\sum\limits_i c_i \ln \frac{M_i}{q}\Bigg)
=\nonumber\\
&& \qquad\qquad\qquad\qquad\quad
= \frac{8\pi^2 n\,q^{2n}/\Lambda^{2n}}{\Big(1+q^{2n}/\Lambda^{2n}\Big)^3}
\sum\limits_i c_i \ln \frac{M_i}{q}
+ \frac{2\pi^2}{\Big(1+q^{2n}/\Lambda^{2n}\Big)^2},\qquad
\end{eqnarray}

\noindent
so that $f(0) = 0$ and $f(\infty) = 2\pi^2$. Then from equation
(\ref{Integrals_DIF}) we obtain, that

\begin{equation}
\Lambda \frac{d}{d\Lambda}
2\pi^2\int d^4q \frac{1}{q^2 (q+p)^2
\Big(1+q^{2n}/\Lambda^{2n}\Big)^2}\sum\limits_i c_i \ln\frac{M_i}{q}
= 4\pi^4 \ln \frac{\Lambda}{p} + O(1)
\end{equation}

\noindent
and, therefore,

\begin{equation}\label{Integrals_LnQ2}
2\pi^2 \int d^4q \frac{1}{q^2 (q+p)^2
\Big(1+q^{2n}/\Lambda^{2n}\Big)^2}\sum\limits_i c_i \ln\frac{M_i}{q}
= 2\pi^4 \ln^2\frac{\Lambda}{p}
+ O\Big(\ln \frac{\Lambda}{p}\Big).
\end{equation}

\noindent
Then it is necessary to calculate logarithmical divergences. For
this purpose we subtract from integral (\ref{Integrals_LnQ2}) terms,
proportional $\ln^2\Lambda/p$ and differentiate the result over
$\ln\Lambda$:

\begin{eqnarray}\label{Integrals_LnQ}
&& \lim\limits_{p\to 0}\,\Lambda\frac{d}{d\Lambda} \Bigg[
2\pi^2 \int d^4q \frac{1}{q^2 (q+p)^2
\Big(1+q^{2n}/\Lambda^{2n}\Big)^2}\sum\limits_i c_i \ln\frac{M_i}{q}
- 2\pi^4 \ln^2\frac{\Lambda}{p} \Bigg]
=\nonumber\\
&& = \lim\limits_{p\to 0}\Bigg\{
- 2\pi^2 \int d^4q \frac{1}{q^2 (q+p)^2}\,q \frac{d}{dq}
\Bigg(\frac{1}{\Big(1+q^{2n}/\Lambda^{2n}\Big)^2}
\sum\limits_i c_i \ln\frac{M_i}{q}\Bigg)
- 4\pi^4 \ln\frac{\Lambda}{p}\Bigg\}
=\nonumber\\
&& = \lim\limits_{p\to 0}\Bigg\{
- 4\pi^4 \int\limits_0^p dq\,\frac{q^2}{p^2}\, \frac{d}{dq}
\Bigg(\frac{1}{\Big(1+q^{2n}/\Lambda^{2n}\Big)^2}
\sum\limits_i c_i \ln\frac{M_i}{q}\Bigg)
-\nonumber\\
&&\qquad\qquad\qquad\qquad\quad
- 4\pi^4 \int\limits_p^\infty dq\, \frac{d}{dq}
\Bigg(\frac{1}{\Big(1+q^{2n}/\Lambda^{2n}\Big)^2}
\sum\limits_i c_i \ln\frac{M}{q}\Bigg)
- 4\pi^4 \ln\frac{\Lambda}{p}\Bigg\}
=\nonumber\\
&&
= 2\pi^4 + 4\pi^4 \sum\limits_i c_i \ln \frac{M_i}{\Lambda}.
\end{eqnarray}

\noindent
From equations (\ref{Integrals_DK_Integral}),
(\ref{Integrals_Atan_Integral}), (\ref{Integrals_LnQ2}) and
(\ref{Integrals_LnQ}) we see, that the divergent part of $I_8$
is equal to

\begin{equation}
I_8 = 2\pi^4\Bigg(\ln^2\frac{\Lambda}{p}
+ 2\ln \frac{\Lambda}{p}\Big(\sum\limits_i c_i \ln \frac{M_i}{\Lambda}
+ \frac{3}{2}\Big)\Bigg) + O(1).
\end{equation}

In order to prove, that integrals $I_9$ and $I_{10}$ are finite at
$\Lambda\to\infty$, first note, that

\begin{equation}
I_9 = I_9\Big(p/\Lambda\Big);\qquad I_{10}=I_{10}\Big(p/\Lambda\Big).
\end{equation}

\noindent
Therefore, it is necessary to prove, that $I_9(p=0)$ and $I_{10}(p=0)$
are finite constants. Let us set $p=0$ and make a substitution

\begin{equation}
k^\mu = \Lambda K^\mu;\qquad q^\mu = \Lambda Q^\mu.
\end{equation}

\noindent
Taking into account, that $M = a\Lambda$, where $a$ is a finite constant,
the considered integrals can be written as

\begin{eqnarray}\label{KQ_I9}
&& I_9 = \int \frac{d^4K}{(2\pi)^4}\frac{d^4Q}{(2\pi)^4}\,
\frac{2 (K+Q)_\mu Q_\mu}{
K^2 \Big(1 + K^{2n}\Big) \Big((K+Q)^2+a^2\Big)^2 \Big(Q^2+a^2\Big)^2};
\qquad\\
\label{KQ_I10}
&& I_{10} = \int \frac{d^4K}{(2\pi)^4}\frac{d^4Q}{(2\pi)^4}\,
\frac{a^2}{K^2 \Big(1 + K^{2n}\Big)
\Big((K+Q)^2+a^2\Big) \Big(Q^2+a^2\Big)^3}.
\end{eqnarray}

\noindent
The integrals over $Q$ are evidently convergent. However, it is necessary
to check, that, after taking these integrals, the remaining integration
over $K$ will be also convergent. Possible divergences can arise at
$K\to 0$ or at $K\to\infty$. In the limit $K\to 0$

\begin{eqnarray}
&& \int \frac{d^4Q}{(2\pi)^4}\,
\frac{2 (K+Q)_\mu Q_\mu}{
\Big((K+Q)^2+a^2\Big)^2 \Big(Q^2+a^2\Big)^2} \to
\int \frac{d^4Q}{(2\pi)^4}\,\frac{2 Q^2}{
(Q^2+a^2)^4} = \frac{1}{24\pi^2 a^2};
\qquad\quad\\
&& \int \frac{d^4Q}{(2\pi)^4}\,
\frac{a^2}{\Big((K+Q)^2+a^2\Big) \Big(Q^2+a^2\Big)^3}\to
\int \frac{d^4Q}{(2\pi)^4}\,
\frac{a^2}{(Q^2+a^2)^4} = \frac{1}{96\pi^2 a^2}.
\end{eqnarray}

\noindent
It means that, in equations (\ref{KQ_I9}) and (\ref{KQ_I10}) the
integration over $K$ is convergent if $K\to 0$. Similarly, in the
limit $K\to\infty$

\begin{eqnarray}
&& \int \frac{d^4Q}{(2\pi)^4}\,
\frac{2 (K+Q)_\mu Q_\mu}{
\Big((K+Q)^2+a^2\Big)^2 \Big(Q^2+a^2\Big)^2} \approx
\int \frac{d^4Q}{(2\pi)^4}\,
\frac{2 (K+Q)_\mu Q_\mu}{(K+Q)^4 Q^4} = \frac{1}{8\pi^2 K^2};
\qquad\quad\\
&& \int \frac{d^4Q}{(2\pi)^4}\,
\frac{a^2}{\Big((K+Q)^2+a^2\Big) \Big(Q^2+a^2\Big)^3} \approx
\int \frac{d^4Q}{(2\pi)^4}\,
\frac{a^2}{(K+Q)^2 (Q^2+a^2)^3}
=\\
&& = \frac{a^2}{8\pi^2}
\int\limits_0^K dQ \frac{Q^3}{K^2 (Q^2+a^2)^3}
+ \frac{a^2}{8\pi^2} \int\limits_K^\infty dQ \frac{Q}{(Q^2+a^2)^3}
= \frac{1}{32\pi^2 (K^2+a^2)} \approx \frac{1}{32\pi^2 K^2},\quad\nonumber
\end{eqnarray}

\noindent
where we used equations (\ref{For_I3}) and (\ref{Angle_Integral}).
Therefore, due to the presence of higher derivative term the integration
over $K$ in equations (\ref{KQ_I9}) and (\ref{KQ_I10}) is also convergent
at $K\to\infty$. Thus the integrals $I_9$ and $I_{10}$ are proven to be
finite in the limit $\Lambda\to\infty$.

Collecting the above results we can finally write integrals
(\ref{Integrals_Definition}) in the following form:

\begin{eqnarray}\label{Integrals_Results}
&& I_1 = 2 \pi^2 \Bigg(\ln\frac{\Lambda}{p} + \frac{1}{2}\Bigg)
+ o(1);\nonumber\\
&& I_2 = 2\pi^2\sum\limits_i c_i \Bigg(\ln\frac{M_i}{p}
+ \sqrt{1+\frac{4M_i^2}{p^2}}\,
\mbox{arctanh}\sqrt{\frac{p^2}{4M_i^2 + p^2}}\Bigg);
\nonumber\\
&& I_3 = 4\pi^4 \ln\frac{\Lambda}{p} + O(1);\vphantom{\Bigg(}\nonumber\\
&& I_4 = \frac{\pi^2}{2} + o(1);\nonumber\\
&& I_5 = 4\pi^4\Bigg(\ln^2 \frac{\Lambda}{p} + \ln\frac{\Lambda}{p}\Bigg)
+ O(1);\nonumber\\
&& I_6 = 4\pi^4\Bigg(\ln^2 \frac{\Lambda}{p} + \ln\frac{\Lambda}{p}\Bigg)
+ O(1);\nonumber\\
&& I_7 = 2\pi^4\Bigg(\ln^2 \frac{\Lambda}{p}
+ 2\ln\frac{\Lambda}{p}\Bigg) + O(1);\nonumber\\
&& I_8 = 2\pi^4\Bigg(\ln^2\frac{\Lambda}{p}
+ 2 \ln \frac{\Lambda}{p}\Big(\sum\limits_i c_i \ln \frac{M_i}{\Lambda}
+ \frac{3}{2}\Big)\Bigg) + O(1);\nonumber\\
&& I_9 = O(1);\vphantom{\frac{1}{2}}\nonumber\\
&& I_{10} = O(1).\vphantom{\frac{1}{2}}
\end{eqnarray}

%%%%%%%%%%%%%%%%%%%%%%%%%%%%%%%%%%%%%%%%%%%%%%%%%%%%%%%%%%%%%%%%%%%%%%%%%%

\pagebreak

\begin{figure}[p]
\hspace*{1.5cm}
\epsfxsize12.0truecm\epsfbox{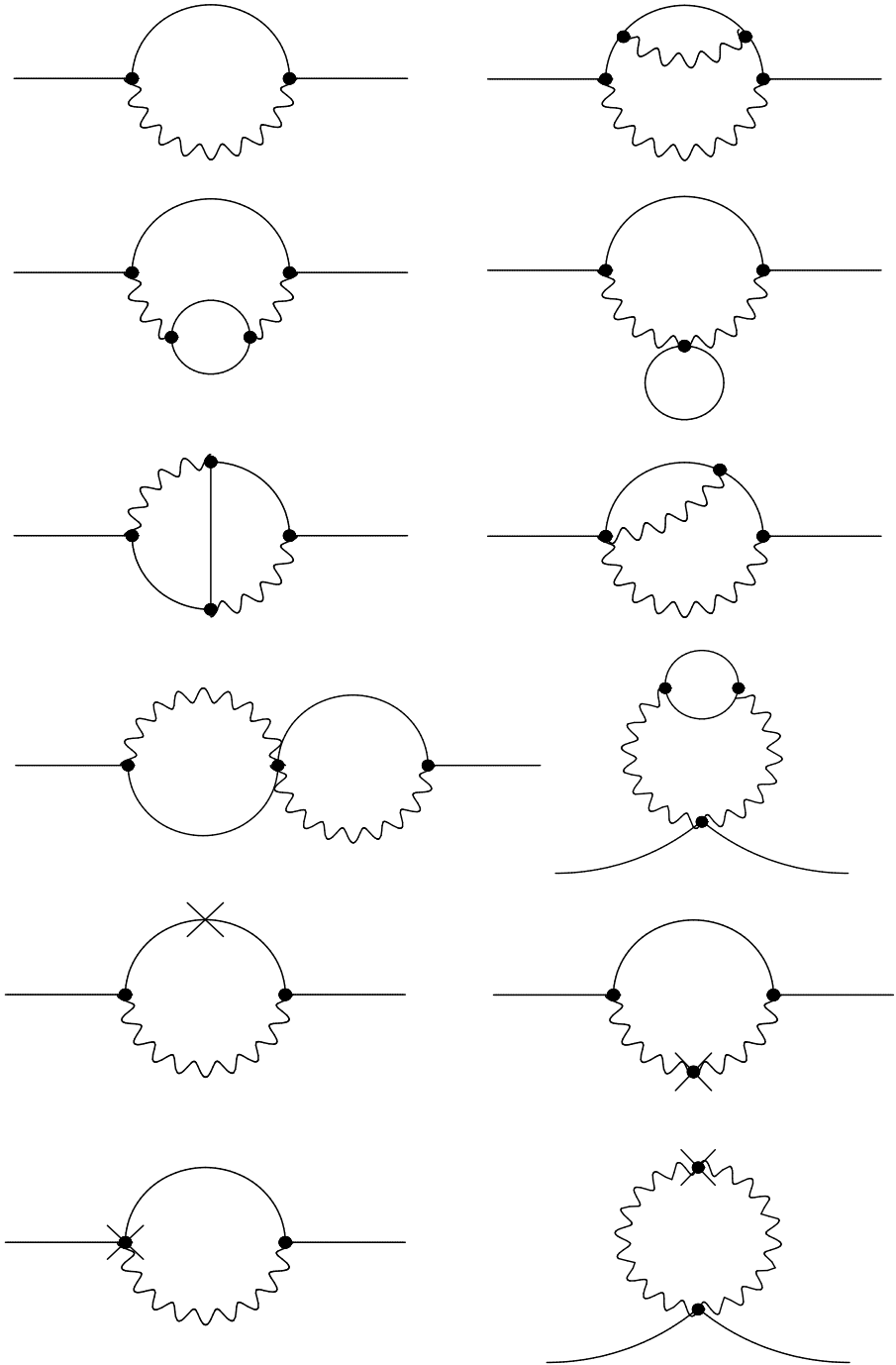}
\caption{Feinman graphs, giving nontrivial contributions to the two-loop
anomalous dimension of $N=1$ supersymmetric electrodynamics.}
\label{Figure_Anomalous_Dimension_Diagrams}
\end{figure}

\end{document}